\documentclass[prr,amsmath,amssymb,twocolumn]{revtex4-2}
\usepackage{graphicx}
\usepackage{subfigure}
\usepackage{adjustbox}
\usepackage{bm}
\usepackage{bbm}
\usepackage{braket}
\usepackage{standalone}
\usepackage{multirow}
\usepackage{tikz}
\usepackage{mathrsfs}
\usepackage{dsfont}
\usepackage[colorlinks,bookmarks=true,citecolor=blue,linkcolor=blue,urlcolor=blue]{hyperref}
\usepackage{cleveref}
\usepackage{comment}
\usepackage{mathtools}
\usepackage{soul}
\usepackage{orcidlink}
\usepackage{color}
\usepackage{cancel}

\definecolor{mzgreen}{RGB}{34, 120, 70}

\numberwithin{equation}{section}
\renewcommand{\theequation}{\arabic{section}.\arabic{equation}}

\def\re{\text{Re}}
\def\im{\text{Im}}
\def\tr{\text{Tr}}

\def\sgn{\text{sgn}}
 \Crefname{equation}{Eq.}{Eqs.}
\Crefname{figure}{Fig.}{Figs.}
\Crefname{section}{Sec.}{Secs.}
\def\tot{\text{tot}}

\def\u{\underline}

\def\diag{\text{diag}}

\begin{document}
\title{Quantum dynamical signatures of non-Hermitian boundary modes}

\author{Fan Yang \orcidlink{0000-0001-8754-7882}}
\affiliation{Department of Physics, Stockholm University, AlbaNova University Center, 10691 Stockholm, Sweden}

\author{Maria Zelenayova \orcidlink{0009-0004-4229-266X}}
\affiliation{Department of Physics, Stockholm University, AlbaNova University Center, 10691 Stockholm, Sweden}

\author{Paolo Molignini \orcidlink{0000-0001-6294-3416}}
\affiliation{Department of Physics, Stockholm University, AlbaNova University Center, 10691 Stockholm, Sweden}

\author{Emil J. Bergholtz \orcidlink{0000-0002-9739-2930}}
\affiliation{Department of Physics, Stockholm University, AlbaNova University Center, 10691 Stockholm, Sweden}

\date{\today}

\begin{abstract}

The non-Hermitian bulk-boundary correspondence features an interplay between the non-Hermitian skin effect and anomalous boundary-mode behavior.
Whereas the skin effect is known to manifest itself in quantum dynamics in the form of chiral damping, it has remained less clear what impact the boundary modes may have. 
Here we derive experimentally accessible signatures of the boundary modes.
We also establish clear criteria, based on the generalized Brillouin zone, that determine when bulk and boundary effects can be dynamically discerned using the Liouvillian separation gap.
This leads to telltale signatures in both stable regimes -- where particle number remains finite -- and in the unstable regimes -- where a macroscopic boundary mode population occurs.
\end{abstract}
\maketitle

\section{Introduction}
In the recent decade, there has been increasing interest in  realizing topological phases of matter in non-Hermitian (NH) systems \cite{ashida2020,emil2021e}. 
Two of the most striking features are the non-Hermitian skin effect (NHSE) \cite{hatano1996,hatano1997,zhang2022r,lee2023} and the breakdown of conventional bulk-boundary correspondence (BBC) \cite{Lee2016,xiong2018,yao2018,flore2018,alvarez2018}. 
The NHSE refers to the localization behaviors of eigenstates towards open boundaries of a lattice system driven by the non-Hermiticity. 
This phenomenon is concurrent with an extreme sensitivity of the spectrum in response to boundary conditions, widely observed across experimental platforms~\cite{helbig2020generalized,Ghatak2020,photonicNHBBC,Brandenbourger_2019,Neupert_2020,ma2022,liang2022d,veenstra2023}.
The spectral sensitivity in 1D and beyond can be captured by the non-Bloch band theory \cite{yao2018,yokomizo2019,wang2024a}, based on which one is able to reconstruct non-Bloch topological invariants and predict the properties of topological \textcolor{black}{NH} boundary modes \cite{yao2018,flore2018,wang2018,yang2024a}. 
NH phenomenology also offers exciting prospects for technological innovations,  including unidirectional single-mode lasing \cite{zhu2022}, topological light funneling \cite{szameit2020}, nonreciprocal state pumping  \cite{Jiangbin2020}, as well as enhanced sensing \cite{Budich2020,mcdonald2020,budich2022q,arandes2024}. 

Going from closed to open quantum systems, 
the NHSE manifests itself as the \emph{Liouvillian skin effect}, prevalent in both non-interacting  and interacting Lindbladians \cite{fei2019,ueda2021,yang2022,kohei2023,yoshida2023,ekman2024,zhong2025m,marco2025,tsuneya2025}. While
dynamical bulk skin modes dominate the relaxations of density matrices under dissipation, the  contribution from NH boundary modes has yet remained elusive. The extent to which NH boundary modes can be dynamically isolated from the bulk, or even  detectable at all in microscopic models, remains unclear~\cite{meng2025}.

\begin{figure}[h]
\centering
\includegraphics[width=1\columnwidth]{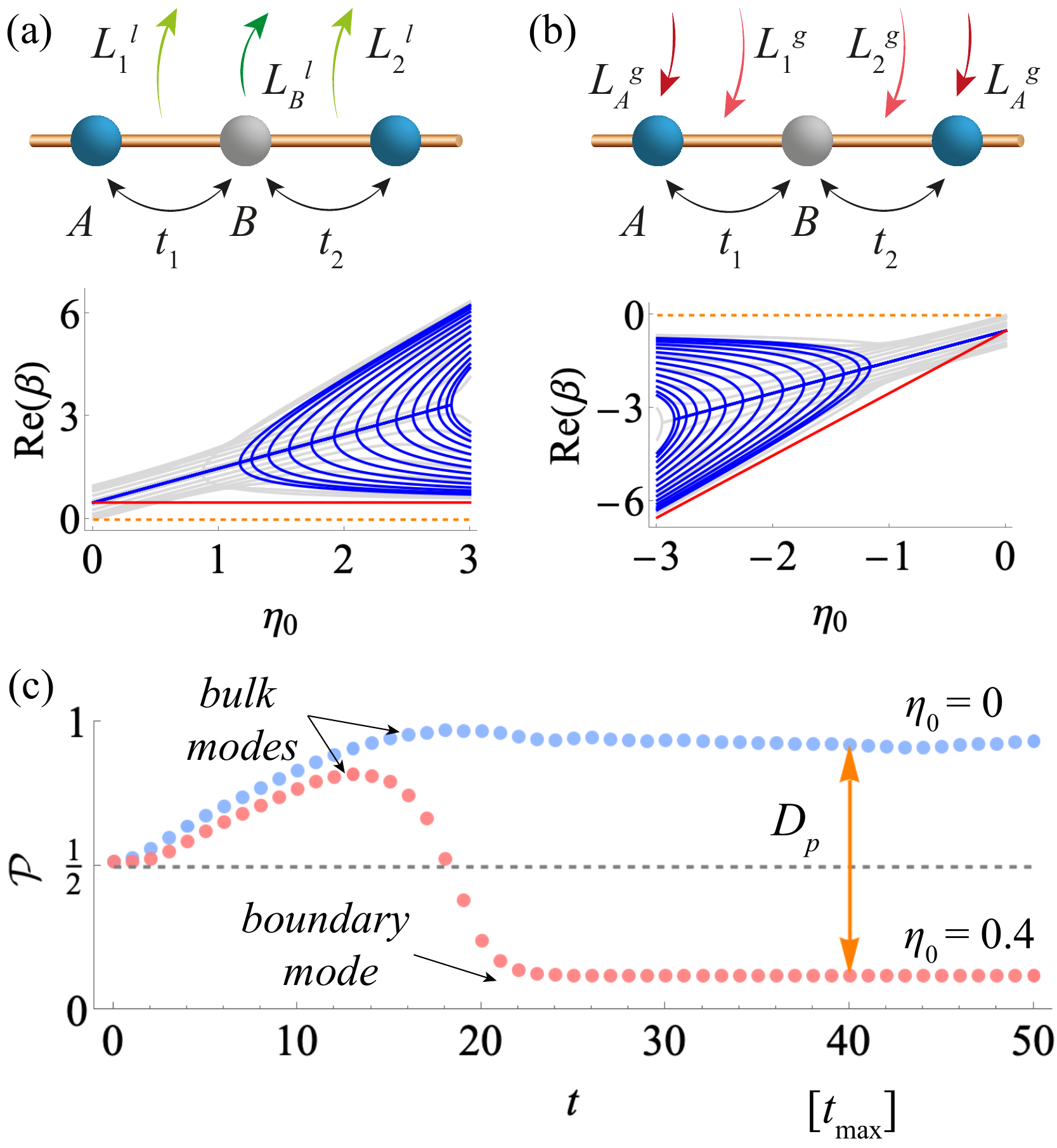} 
\caption{{\bf Distillation vs. Amplification: two scenarios to isolate NH boundary mode in bosonic Lindbladians.} 
(a) Distillation scheme with loss on bonds and sublattice $B$. 
(b) Amplification scheme with gain on bonds and sublattice $A$.
The bottom panels depict the real part of the rapidity spectrum (OBC), shown as a function of sublattice dissipation $\eta_0$ for $N_{\text{tot}} = 2N-1=29$ sites at fixed (a) $t_1 =1, t_2=2, \eta_1=0.5, \eta_2=0$ and (b) $t_1 =-1, t_2=2, \eta_1=-0.5, \eta_2=0$. 
The Liouvillian gap between boundary (red) and bulk (blue) modes is enlarged by increasing $|\eta_0|$. 
The PBC rapidity spectrum is shown in gray. 
(c) Illustration of polarization drift $D_p$ in \Cref{eq:dyp} as a change in long-time polarization when increasing $\eta_0$ from  $0$ to $0.4$ in the distillation scenario (a). From the density in \Cref{eq:ntb} and \Cref{eq:nts}, NH bulk and boundary modes contribute to distinct localization tendencies of bosons: $|r|^{-1} \simeq 1.73 > 1$ and $ |r_L| = 0.75 < 1$, resulting in  $\mathcal{P} (t_\text{max})\to 1$   and $\mathcal{P}(t_\text{max}) \to 0$, respectively.
}
\label{fig:model}
\end{figure}

Here, we tackle this question by studying a class of exactly solvable quadratic bosonic Lindbladians (QBLs). We show that when the system exhibits a positive \emph{Liouvillian separation gap} --  defined from the gap on the real Liouvillian spectrum separating a boundary mode from all the bulk modes  -- the boundary mode becomes dynamically dominant (\Cref{fig:model}). Importantly, we find that given comparatively weak nonreciprocity in the non-Bloch Hamiltonians, such NH boundary-mode separation  is generically possible by introducing sublattice dissipation (\Cref{fig:sg}).
Engineered dissipation has been known to  help select topologically protected Hermitian midgap and boundary states
\cite{schomerus2013,emil2023,meng2023,fei2023s,hu2024,ma2025}. We now extend its application to the NH regime.

On the other hand, compared with open Fermi matter \cite{prosen2008,prosen2010ex,prosen2010sp,eisert2010,alba2021,barthel2022,clerk2023t}, the dynamics of open bosons remains less explored, with previous studies focusing on steady-state properties~\cite{clerk2022,zhang2022c,zhong2023s}.
One of the challenges lies in the unbounded bosonic Liouville space \cite{prosen2010q}, giving rise to dynamical instabilities in finite-size systems with open boundaries \cite{viola2024}. Nevertheless, 
such instabilities open the door to unique bosonic phenomena such as directional amplification~\cite{wanjura2020} and NH lasing~\cite{Feng_NH_laser_Science_2014, NH_laser_mercedeh_2018}. 
The instability of bosonic Lindbladians occurs when the sign of the \emph{Liouvillian gap} \cite{prosen2010q,viola2024} -- defined as the minimum of the entire Liouvillian real spectrum -- turns negative, which implies a positive growth rate in boson densities.  In our models, it pinpoints the unstable regime when global gain surpasses loss [\Cref{fig:model}~(b)].
Moreover, akin to quadratic fermionic Lindbladians \cite{cooper2020}, we find that the topology of QBLs is not directly linked to steady states, but encoded in the NH damping matrix, which determines the relaxation dynamics of the system and can be mapped to the non-Bloch Hamiltonian of our interest. Based on this, we
trace the underlying NH BBC through the dynamical measurement of \emph{polarization drift}, the change in long-time polarization with and without the boundary-mode separation [\Cref{fig:model}~(c) and \Cref{fig:drift}].


In a full open quantum framework, our work characterizes the dynamical separability 
of NH boundary modes. A successful separation facilitated by engineered  dissipation not only sheds light on the NH topology of Lindbladians, but  also offers the prospect of future design of NH photonic, atomic and  superconducting platforms where boundary-mode engineering is critical. 

The paper is structured as follows: first, we introduce a general scheme to modulate Liouvillian separation gap with sublattice dissipation  (\Cref{sec:gf}). Specifically, we employ QBLs and present a minimal model with exact solutions, the bond-dissipative Su-Schrieffer-Heeger (SSH) chain with odd sites (\Cref{sec:model}).  Second, by probing boson densities, we witness dynamical signatures of NH boundary modes in terms of 
\textcolor{black}{distinct damping behaviors}  and polarization drift. We relate these signatures to quantum phase transitions and unveil the NH BBC  of  the non-Bloch Hamiltonian (\Cref{sec:dyn}). Third, we address the instabilities and singularities in our QBLs, accompanied by an analysis on the asymptotically unstable finite-size dynamics (\Cref{sec:is}).

\section{General scheme} 
\label{sec:gf}

To begin with, we introduce the Liouvillian separation gap in QBLs and relate its sign to the  dynamical separability of NH boundary modes. To effectively modulate the separation gap, a general approach is formulated based on engineered sublattice dissipation.

\subsection{Liouvillian separation gap}
\label{sec:lsg}
The dynamics of an open quantum system is described by the Lindblad master equation \cite{lindblad1976,breuer2007}:
\begin{gather}
  \frac{d \rho}{dt} = \hat{\mathcal{L}} \rho \coloneqq -i [ \mathcal{H}, \rho ] + \sum_{\mu} ( L_\mu \rho L_\mu^\dagger - \frac{1}{2}\{ L_\mu^\dagger L_\mu, \rho \} ),
  \label{eq:lin}
\end{gather}
where $\rho$ denotes the density matrix and $\mu$  a sum over all types of Lindblad dissipators.
We consider a lattice model of $n$ sites captured by a tight-binding Hamiltonian $\mathcal{H} = \u{b}^\dagger \cdot H \u{b}$. Here, we adopt the notation for a column vector  of scalars or operators: $\u{x} = (x_1, x_2, \dots, x_n)^T$ and $\u{x}(j) = x_j$.  $b_j^\dagger$ ($b_j$) creates (annihilates) a boson on the $j$-th site. 
Loss and gain are encoded in linear Lindblad jump operators: $L_\mu^l = \u{l}_\mu \cdot \u{b}$  and $L_\mu^g = \u{g}_\mu \cdot \u{b}^\dagger$.

Applying third quantization for bosons \cite{prosen2010q}, we verify in the Appendix that the eigenvalues (or \emph{rapidities}) of the Liouvillian coincide with those of the NH damping matrix:
\begin{gather}
    X = iH^T + \frac{1}{2} \left[ (M^l)^T - M^g \right], \label{eq:x}
\end{gather}
where $M^l_{ij} = \sum_\mu \u{l}_{\mu}^*(i) \u{l}_{\mu}(j)$ and $M_{ij}^g = \sum_\mu \u{g}_\mu^*(i) \u{g}_\mu(j)$ with $*$ denoting the complex conjugation. 
Any physical observable of QBLs can be obtained from the two-point correlator $C_{jk}(t) = \tr[\rho(t) b_j^\dagger b_k]$ by the Wick theorem. Its equation of motion \cite{eisert2010,clerk2023t} obeys 
\begin{gather}
    \partial_t C(t) = -X^\dagger C(t) - C(t) X + M^g. \label{eq:eom}
\end{gather}
For completeness, we re-derive \Cref{eq:eom} in the Appendix using commutation relations of bosons: $[b_i, b_j^\dagger] = \delta_{ij}$. 

The correlation function is solvable if there exists a covariance matrix $C_s$ that satisfies $\partial_t C_s=0$, or equivalently,
\begin{gather}
   X^\dagger C_s + C_s X = M^g. \label{eq:rss0}
\end{gather}
The non-existence of $C_s$  signifies a singularity in QBLs, the discussion of which we leave to \Cref{sec:is}.
For the moment, we assume $C_s$ is known. Taking  $C_s$ as a reference point in the correlator $\tilde{C}(t) = C(t) - C_s$,
the equation of motion can be recast into 
\begin{gather}
   \partial_t \tilde{C}(t) =  -X^\dagger \tilde{C}(t) - \tilde{C}(t)X. 
\end{gather}
Given an initial state at $t=0$,  $ \tilde{C}(t) = e^{-X^\dagger t} \tilde{C}(0) e^{-X t} $.  We then apply the biorthogonal eigenvalue decomposition \cite{brody2013} of the NH damping matrix:
\begin{gather}
    X = \sum_m \beta_m |\u{\psi}_{R,m}\rangle \langle \u{\psi}_{L,m}|, \notag \\ \langle \u{\psi}_{L,m}  | \u{\psi}_{R,m'} \rangle = \delta_{m,m'}, \label{eq:xdom}
\end{gather}
and arrive at a compact closed form of the two-point correlator:
\begin{gather}
\tilde{C}(t) = \sum_{m,m'} e^{-(\beta_m^* + \beta_{m'})t} |\u{\psi}_{L,m}\rangle I_{m,m'} \langle \u{\psi}_{L,m'}|. \label{eq:tc}
\end{gather}
Here, $I_{m,m'} = \langle \u{\psi}_{R,m} | \tilde{C}(0) | \u{\psi}_{R,m'} \rangle$ and  $m$ denotes the band index.  

For each eigenmode $m$ in $\tilde{C}(t)$, the imaginary part of rapidity ($\im[\beta_m]$) governs the phase oscillations and the real part determines its decay ($\re[\beta_m] > 0$) or growth ($\re[\beta_m] < 0$) rate. From the sign of the Liouvillian gap \cite{minganti2018,ueda2021}:
 \begin{gather}
    \Delta \equiv 2 \min_{\forall m} \re[\beta_m],  \label{eq:lg}
 \end{gather}
we can further tell whether the QBLs are stable or not.
When $\Delta > 0$, 
$\tilde{C}(\infty)  = C(\infty) - C_s = 0$. At long times, the system relaxes to the steady state, the correlation function of which is given by $C_s$.  Whereas, when $\Delta < 0$, the boson number ${n}_j(t) = C_{jj}(t)$ diverges:  $ |\tilde{n}_j(\infty)|  = |n_j(\infty) - n_{j,s}| = \infty$. The system becomes dynamically unstable beyond short times. We thus refer to the state described by 
$C_s$ as the bosonic reference steady state (BRSS). It becomes a real steady state only when QBLs hold a positive Liouvillian gap.

Now, we map the damping matrix in  \Cref{eq:x} to a NH Hamiltonian, the topology of which is our main interest: 
\begin{gather}
    X = \text{const} \cdot \mathbbm{1} + i\mathcal{H}_{\text{NH}}. \label{eq:xh}
\end{gather}
\textcolor{black}{This mapping is generic with the}
non-Hermiticity in $\mathcal{H}_{\text{NH}}$ coming from the \textcolor{black}{Hermitian} loss and gain matrices ($M^l$, $M^g$). In particular,  when correlated bond dissipation is introduced via reservoir engineering \cite{clerk2015n,clerk2022i,he2022,begg2024q}, $\mathcal{H}_{\text{NH}}$ becomes a  nonreciprocal Hamiltonian~\cite{fei2019,yang2022} (see also our example in \Cref{sec:set}).

Since $X$ and $\mathcal{H}_{\text{NH}}$ share the same set of eigenmodes $\{ m \}$, at short times the boundary ($m=0$) and bulk ($m\neq0$) modes of $\mathcal{H}_{\text{NH}}$ contribute simultaneously to $\tilde{C}(t)$.  After a long-time evolution, however, it is the mode with a smaller Liouvillian gap that dominates. 
We are ready to introduce a key diagnostic of the dynamical separability of NH boundary modes, the Liouvillian separation gap:
\begin{gather}
\Delta_s \equiv \Delta_{\text{bulk}} - \Delta_{\text{boundary}}. \label{eq:sgap}
\end{gather}
It is defined as the difference between the Liouvillian gaps of bulk and boundary modes in \Cref{eq:lg}. More precisely,
\begin{gather}
    \Delta_{\text{bulk}} = 2 \min_{\forall m \ne 0} \re[\beta_m], \notag \\
    \Delta_{\text{boundary}} = 2 \beta_0. \label{eq:lgbb}
\end{gather}
A positive (negative) separation gap $\Delta_s > 0$ ($\Delta_s < 0$) implies that  the long-time dynamics is governed by the boundary (bulk) modes. Illustrated in the real rapidity spectrum of \Cref{fig:model}(a) and (b), when $\Delta_s > 0$ with the boundary mode (in red) lying below all the bulk modes (in blue), the boundary mode appears as the slowest decaying ($\re[\beta_0] > 0$) or the fastest growing ($\re[\beta_0] < 0$) mode during relaxations [\Cref{fig:bm}~(b),(c),(e) and (f)].  Apart from $\Delta_s < 0$, $\Delta_s = 0$ also precludes an effective separation of the boundary mode [\Cref{fig:bm}~(a) and (d)] as the number of bulk modes prevails in the spectrum or the spectral vicinity.

\subsection{Gap engineering with sublattice dissipation}
\label{sec:ge}
Without any modulation, the Liouvillian separation gap is in general non-positive. In the simple example, we take $\mathcal{H}_{\text{NH}}$ as a nonreciprocal Hamiltonian  on a one-dimensional (1D) lattice. To analyze Liouvillian separation gap, one relates the rapidities to eigen energies $\epsilon_m$ of $\mathcal{H}_{\text{NH}}$ from the direct mapping in \Cref{eq:xh}:
 \begin{gather}
     \beta_m = \text{const} + i \epsilon_m.
 \end{gather}
Under the open boundary condition (OBC), $\epsilon^\text{bulk}_m$ in the thermodynamic limit converges to the spectrum of a non-Bloch Hamiltonian, which is associated to its Bloch counterpart by an imaginary momentum shift~\cite{yao2018,yokomizo2019,flore2019} 
  \begin{gather}
      H_{\text{NH}}^{\text{non-Bloch}}(k)  =   H_{\text{NH}}^{\text{Bloch}}(k - i\ln r). \label{eq:gbz}
  \end{gather}
  After the shift, the module of the Bloch phase factor $|e^{i(k-i\ln r)}| = |r|$ deviates from $1$, a feature characteristic of the generalized Brillouin zone (GBZ). 
Suppose $\mathcal{H}_{\text{NH}}$ hosts a boundary mode at zero energy $\epsilon_0 = 0$. The separation gap is simplified to 
 \begin{gather}
     \Delta_s =  \min_{\forall m \ne 0} \{ -2 \im[\epsilon_m]\}.
 \end{gather}
Once there exists a bulk eigenmode $m$ such that $\im[\epsilon_m] \ge 0$, we have $\Delta_s \le 0$, indicating that bulk modes dominate the dynamics over the boundary mode. Most generic NH Hamiltonians fall into this category. 

Furthermore,  we can impose a $P$ symmetry \cite{bernard2002,bernard2020} on 
the non-Bloch Hamiltonian, which is equivalent to  chiral symmetry in case of Hermitian Hamiltonians:
\begin{gather}
H_{\text{NH}}^{\text{non-Bloch}}(k) = -P H_{\text{NH}}^{\text{non-Bloch}}(k) P^{-1}, \  P^2 = \mathbbm{1}.  \label{eq:chiral}
\end{gather}
Bulk eigen energies then come in opposite pairs: $\{ \epsilon_m \} = \{ -\epsilon_m \}$. It immediately leads to $\Delta_s \le 0$. 

To flip the sign of $\Delta_s$ or modulate the separation gap, we introduce sublattice dissipation in accordance with the spatial structure of the targeted boundary mode \cite{emil2023}. Any NH boundary mode can be approximated by the following ideal structure: 
it is exponentially localized on the sublattice $A$ and vanishes completely on other sublattices $\bar{A}$. To enhance its dynamical contribution, one can employ two intuitive approaches: applying $\bar{A}$-sublattice loss with $L_{0,j}^l = \sqrt{\gamma_0^l} b_j$,  $\forall j \in \bar{A}$ or adding $A$-sublattice gain with $L_{0,j}^g = \sqrt{\gamma_0^g} b_j^\dagger$, $\forall j \in A$. They correspond to our distillation and amplification schemes [\Cref{fig:model}(a) and (b)]. The uniformity of sublattice dissipation brings maximal modulation effects on $\Delta_s$ and at the same time, ensures that the boundary mode remains an eigenmode of $\mathcal{H}_{\text{NH}}$.

To elucidate our protocol, we continue with 1D $P$-symmetric model and enrich it with two bands: 
\begin{gather}
    H_{\text{NH}}^{\text{non-Bloch}}(k) = d_x(k)\sigma_x+d_y(k)\sigma_y. \label{eq:hch}
\end{gather} 
 Here, the components of two Pauli matrices $\sigma_{\alpha = x,y}$ are complex: $d_\alpha = d_{\alpha, R} + id_{\alpha, I}$ with $d_{\alpha, R/I} \in \mathbb{R}$. The $P$ symmetry of \Cref{eq:chiral} can be represented by a matrix $P = \diag \{1, -1 \}$, which leads to $\Delta_s \le 0$ in absence of modulation. 
Adding modulation, the damping matrix  in \Cref{eq:x} and \Cref{eq:xh} takes a new form:  $X \to X+\Delta X$,
 \begin{gather}
   \Delta X = \eta_0 \cdot \mathbbm{1} + id_z \sigma_z, \quad  d_z = i |\eta_0|.
   \label{eq:subl}
 \end{gather}
The strength of sublattice dissipation is denoted by  $\eta_0 = \frac{1}{4}|\gamma_0^l|$ and $\eta_0 = -\frac{1}{4}|\gamma_0^g|$  in the distillation and amplification scenarios, respectively.  As a result, the  rapidity spectra are changed into $\beta_\pm (k) = \text{const} + \eta_0 \pm i \sqrt{  d_x^2(k) + d_y^2(k) - \eta_0^2}$ for bulk modes, and $\beta_0 = \text{const} + \eta_0 - |\eta_0|$ for the boundary mode. Liouvillian separation gap becomes a function of $\eta_0$:
\begin{gather}
\Delta_s = \min_{k} 2 \re f(k), \label{eq:deltas}\\
f(k) = |\eta_0| - \sqrt{\eta_0^2 - \sum_{\alpha=x,y} (d_{\alpha,R}^2 - d_{\alpha,I}^2 + 2id_{\alpha,R}d_{\alpha,I}) }. \notag
\end{gather}
To determine the sign of $\Delta_s$, we first look at two extremes of $\eta_0$. At $|\eta_0| =  0$, $\Delta_s \le 0$ from the $P$ symmetry. When $|\eta_0| \to \infty$, $\Delta_s$ approaches zero. In both cases, the sign of $\Delta_s$ cannot be effectively flipped. Whereas, for finite $\eta_0$ on a scale $\mathcal{O}(|\eta_0|) \sim \max_{\forall \alpha}\{\mathcal{O}(|d_{\alpha,R}|), \mathcal{O}(|d_{\alpha,I}|) \}$, it is easy to discern that  
\begin{align}
    &f(k) \simeq  \\
  &  \begin{cases}
        |\eta_0| - \sqrt{\eta_0^2 - \sum_{\alpha}d_{\alpha,R}^2}, & \text{if} \ \textcolor{black}{\mathcal{O}(|d_{\alpha,I}|) <  \mathcal{O}(|d_{\alpha,R}|)} \\
        |\eta_0| - \sqrt{\eta_0^2 + \sum_{\alpha}d_{\alpha,I}^2}, & \text{if} \ \textcolor{black}{\mathcal{O}(|d_{\alpha,I}|) > \mathcal{O}(|d_{\alpha,R}|)} 
    \end{cases}. \notag
\end{align}
The sign difference in the square comes from 
the nonreciprocity (non-Hermiticity ) in $id_{\alpha,I}$. While nonreciprocity hinders the formation of  a positive $\Delta_s$, reciprocity reinforces it. We thus arrive at one of the main results of our work:
 \begin{gather}
  \textcolor{black}{\Delta_s > 0, \quad \text{if} \quad \mathcal{O}(|d_{\alpha,I}|) <  \mathcal{O}(|d_{\alpha,R}|), \quad  |\eta_0| \ne 0.} \label{eq:main}
 \end{gather} 
It shows that the dynamical separation of NH boundary mode is generically possible through engineered dissipation when the nonreciprocity is not dominant. Particularly, as $|d_{\alpha,I}| \to 0$,  $\mathcal{H}_{\text{NH}}$ regresses to a Hermitian Hamiltonian for which $\Delta_s$ stays positive at infinitesimal $\eta_0$. It guarantees an effective dissipative preparation of Hermitian boundary modes \cite{emil2023,meng2023}. 
In the ensuing more specific NH models (\Cref{fig:model}), we will see that in fact, the region of weak nonreciprocity
 accommodates a large parameter space (\Cref{fig:sg}),  exhibiting a Liouvillian skin effect (\Cref{fig:bm}) and allowing the witness of quantum phase transitions (\Cref{fig:drift}).

\section{The model}
\label{sec:model}
As a case study, we present an exactly solvable model, the NH bosonic SSH chain with odd sites. Its non-Bloch Hamiltonian shares the $P$-symmetric form in \Cref{eq:hch}. We introduce the nonreciprocity through bond dissipation and apply our distillation and amplification schemes targeting on its NH boundary mode. Based on exact Liouvillian spectra, a comprehensive analysis of the separation gap is carried out in the full range of nonreciprocity.

\subsection{Symmetries, exact boundary mode and biorthogonal polarization}
\label{sec:sym}
We are interested in the generalized NH bosonic SSH chain with an odd number of sites~\cite{elisabet2020}:
\begin{align}
    \mathcal{H}_{\text{NH}} \coloneqq  \mathcal{H}_{\text{S}} =\sum_{j=1}^{N-1}  &t_1^+ b_{j,A}^\dagger b_{j,B} 
    + t_1^- b_{j,B}^\dagger b_{j,A} \notag \\
    + &t_2^+ b_{j,B}^\dagger b_{j+1,A} + t_2^- b_{j+1,A}^\dagger b_{j,B}.
\end{align}
For simplicity, we  adopt
real nonreciprocal hopping parameters: $t_i^\pm \in \mathbb{R}$.
It enforces a $K$ symmetry \cite{bernard2002,bernard2020} on  $\mathcal{H}_{\text{S}}$:
\begin{gather}
   \mathcal{H}_{\text{S}} = K \mathcal{H}_{\text{S}}^* K^{-1}, \quad KK^* = \pm \mathbbm{1}. \label{eq:ksym}
\end{gather}
Eigen energies of $\mathcal{H}_\text{S}$  come in complex conjugate pairs: $\{ \epsilon_m \} =  \{ \epsilon_m^{*} \}$.
Representing $\mathcal{H}_{\text{S}}$ as a real-space matrix with   $n = n_\text{tot} = 2N-1$ sites,
\begin{gather}
    H_\text{S} =   \begin{pmatrix}
           0 & t_1^+ & &   \\
           t_1^- & 0 & t_2^+   &   \\
            & t_2^- & 0  &   \\        
            & &  &  \ddots 
         \end{pmatrix}, \label{eq:hs0}
\end{gather}
the $K$ symmetry is identified as $K = \mathbbm{1}_{n \times n}$. The
$P$ symmetry in \Cref{eq:chiral} also extends from the momentum space to the real space: 
 \begin{gather}
H_\text{S} = -P H_\text{S} P^{-1}, \quad P^2 = \mathbbm{1}, \label{eq:psym}
\end{gather}
with a matrix representation,
\begin{gather}
     P =  \begin{pmatrix}
       1 & & &   \\
           & -1 &   &    \\
            & & 1  &   \\        
            & &  &  \ddots 
     \end{pmatrix}. \label{eq:ps}
 \end{gather}
Eigen energies come in opposite  pairs: $\{ \epsilon_m \} = \{ -\epsilon_m \}$. It follows that given an odd number of eigenmodes, $\mathcal{H}_\text{S}$ must have at least one  eigenmode at $\epsilon_0 = 0$, which turns out to be the boundary mode.

From the eigenvalue equations:
\begin{gather}
    H_\text{S} \  {\u{\psi}}_{Rm} = \epsilon_m {\u{\psi}}_{Rm}, \quad
    H_\text{S}^\dagger \ {\u{\psi}}_{Lm}  = \epsilon_m^* {\u{\psi}}_{Lm},
\end{gather}
we arrive at an exact analytical structure for this zero-energy boundary mode~\cite{flore2018,yang2022}:
 \begin{gather}
  \u{{\psi}}_{R/L,0}^T = \mathcal{N}_{R/L}  \begin{pmatrix}
    r_{R/L} & 0 & r^2_{R/L} & \dots  & 0  & r^N_{R/L}  
  \end{pmatrix}.
  \label{eq:bmm}
 \end{gather}
The localization parameters read
\begin{gather}
    	 r_{R} = -\frac{t_1^- }{t_2^+}, \quad r^*_{L} = -\frac{t_1^+}{t_2^-}. \label{eq:rRL}
    \end{gather}
We also introduce normalization factors $\mathcal{N}_{R/L}$  to meet the biorthogonal relation $ \langle {\u{\psi}}_{L,0}  |\u{{\psi}}_{R,0} \rangle = 1$:
    \begin{align}
   & \mathcal{N}^*_L\mathcal{N}_R = \\
   & \begin{cases}
\  (r^*_{L}r_{R}-1)/\{(r^*_{L}r_{R})[ (r^*_{L}r_{R})^{N}-1] \},  & |r^*_{L}r_{R}| \ne 1; \\
   \  N^{-1},    & \ r^*_{L}r_{R} =  1; \\
  \  [(-1)^N - 1]/2,  & \  r^*_{L}r_{R} =  -1.
    \end{cases} \notag
   \end{align}
   
The phase transitions of the NH SSH chain under OBC are captured by the 
biorthogonal polarization $\mathcal{P}_0$ of the boundary mode \cite{flore2018,elisabet2020,yang2022,yang2024a}:
 \begin{gather}
    \mathcal{P}_0 =  1- \lim_{N \to \infty} \frac{1}{N} | \langle \u{\psi}_{L,0}| \sum_{j=1}^N j \Pi_j | \u{\psi}_{R,0}\rangle |. \label{eq:pol}
 \end{gather}
The density operator  $\Pi_j = \sum_{\lambda} |j, \lambda \rangle \langle j,\lambda|$ with $|j, \lambda \rangle = b^\dagger_{j,\lambda} |0\rangle$ and $\lambda=A,B$ measures the boson density in the $j$-th unit cell.
Taking into account the localization profile of the boundary mode, it is clear that 
 \begin{align}
     \mathcal{P}_0 = 
     \begin{cases}
         1, \   &|r_L^*r_R| < 1 \\
        0, \   &|r_L^*r_R| > 1
     \end{cases}. \label{eq:jump}
 \end{align}
 $\mathcal{P}_0$ exhibits a quantized jump  at
 \begin{gather}
     |r_L^*r_R| = 1 \quad \Leftrightarrow \quad |t_1^2 - \eta_1^2| = |t_2^2 - \eta_2^2|, \label{eq:pt}
 \end{gather}
which reflects the NH BBC in the sense that these lines locate gapless phase transitions where the boundary mode enters the bulk spectrum. 

So far, $\mathcal{P}_0$ is defined as the polarization of a pure state in closed systems. We will later generalize it to the dynamical polarization of a  mixed state [\Cref{eq:dpol}] in our attempt to  probe the NH BBC in open systems (\Cref{sec:pol}).

\subsection{Setup}
\label{sec:set}

From \Cref{eq:bmm}, the boundary mode of  $\mathcal{H}_\text{S}$   is exponentially localized on the sublattice $A$ and fully suppressed on the sublattice $B$, fulfilling the 
structure in our separation schemes (\Cref{sec:ge}).
Next, we detail the general schemes by building our model in an open quantum setup composed of dissipative free bosons.

First, to engineer $\mathcal{H}_{\text{S}}$ we begin with a bosonic Hermitian SSH chain: 
\begin{gather}
    \mathcal{H} = \sum_{j=1}^{N-1} t_1 b_{j,A}^\dagger b_{j,B} + t_2 b_{j+1,A}^\dagger b_{j,B} + {\it h.c.}. \label{eq:her}
\end{gather}
Following the Lindblad master equation, 
nonreciprocity can be realized in the damping matrix  by introducing bath coherent loss and gain on $t_1$ and $t_2$ bonds:
\begin{align}
  &\begin{cases}
      L^l_{1,j} &= \sqrt{\gamma^l_{1}} (b_{j,A} - i  b_{j,B})  \\
      L^g_{1,j} &= \sqrt{\gamma^g_{1}} (b_{j,A}^\dagger + i b_{j,B}^\dagger)
   \end{cases}, \notag \\
  &\begin{cases}
     L^l_{2,j} &= \sqrt{\gamma^l_{2}} (b_{j,B} - ib_{j+1,A})  \\
     L^g_{2,j} &= \sqrt{\gamma^g_{2}} (b_{j,B}^\dagger + ib_{j+1,A}^\dagger)
  \end{cases}. \label{eq:bs}
  \end{align}
  It is important to adapt bond dissipation to the boundary condition. 
For an OBC chain with $(2N-1)$ sites, we start from a chain with $2N$ sites under the periodic boundary condition (PBC) and take away  $B$ site in the last unit cell. In the bulk ($j=1, \dots, N-1$), the bond dissipation in \Cref{eq:bs} remains intact. At the end of the chain ($j=N$),  only on-site dissipation survives: 
\begin{align}
   &\begin{cases}
      L^l_{1,N} &= \sqrt{\gamma^l_{1}} b_{N,A}  \\
      L^g_{1,N} &= \sqrt{\gamma^g_{1}} b_{N,A}^\dagger
   \end{cases}, \notag \\
   &\begin{cases}
     L^l_{2,N} &= (-i)\sqrt{\gamma^l_{2}} b_{1,A}  \\
     L^g_{2,N} &= i\sqrt{\gamma^g_{2}}  b_{1,A}^\dagger
  \end{cases}, \label{eq:bsb}
\end{align}
where we discard $b_{N,B}, b^\dagger_{N,B}$ in \Cref{eq:bs} and recognize identical sites $(N+1,A) = (1,A)$.

The damping matrix in \Cref{eq:x} is then built with bond loss and gain matrices,
\begin{align}
    (M_b^l)^T &= (|\gamma_1^l| +|\gamma_2^l|) \cdot \mathbbm{1} + i \begin{pmatrix}
           0 & |\gamma_1^l| & &   \\
           -|\gamma_1^l| & 0 & |\gamma_2^l|  &    \\
            & -|\gamma_2^l| & 0  &   \\        
            & & &   \ddots 
         \end{pmatrix}, \notag \\
    M_b^g &= (|\gamma_1^g| +|\gamma_2^g|) \cdot \mathbbm{1}  + i \begin{pmatrix}
           0 & |\gamma_1^g| & &   \\
           -|\gamma_1^g| & 0 & |\gamma_2^g|  &    \\
            & -|\gamma_2^g| & 0  &   \\        
            & & &   \ddots 
         \end{pmatrix}. \label{eq:mg}
\end{align}
It leads to 
 \begin{gather}
     X = (\eta_1 + \eta_2) \cdot \mathbbm{1} + i H_\text{S}.
     \label{eq:damp0}
 \end{gather}
To match $H_\text{S}$ in \Cref{eq:hs0}, we identify  ($i = 1,2$)
 \begin{gather}
    t_i^\pm = t_i \pm \eta_i, \notag \\
    \eta_i = \frac{1}{2} (|\gamma^l_i|-|\gamma^g_i|). \label{eq:etai}
 \end{gather}
With bosons, the asymmetric hopping parameters $\eta_i$ are generated by the imbalance between bond loss and gain.  Whereas, with fermions, the asymmetric terms are equal to the sum of bond loss and gain \cite{fei2019,yang2022}.

 According to our separation schemes, we can enhance the dynamical contribution from the boundary mode of $\mathcal{H}_\text{S}$ in \Cref{eq:bmm} by adding 
 uniform loss on the sublattice $B$: 
  \begin{gather}
\forall j: \quad  L^l_{0,j}  = \sqrt{\gamma^l_{0}} b_{j,B}, \label{eq:sdb}
\end{gather}
or uniform gain on the sublattice $A$:
  \begin{gather}
\forall j: \quad  L^g_{0,j}  = \sqrt{\gamma^g_{0}} b^\dagger_{j,A}. \label{eq:sda}
\end{gather}
Sublattice loss and gain matrices take the form:
\begin{align}
    (M_0^l)^T &= \begin{pmatrix}
           0 &  & &   \\
           & |\gamma_0^l| &  &    \\
            &  & 0  &   \\        
            & & &   \ddots 
         \end{pmatrix} = \frac{|\gamma_0^l|}{2} \left[\mathbbm{1} + i \cdot (iP) \right], \\
    M_0^g &= \begin{pmatrix}
           |\gamma_0^g| &  & &   \\
           & 0 &  &    \\
            &  & |\gamma_0^g|  &   \\        
            & & &   \ddots 
         \end{pmatrix}  = \frac{|\gamma_0^g|}{2} \left[\mathbbm{1} + i \cdot (-iP)\right],
         \notag
\end{align}
where the matrix $P$ is employed from \Cref{eq:ps}. 

It is convenient to apply a pure bath such that sublattice loss (gain) is paired with bond loss (gain), which we refer to as the distillation (amplification) scenario in \Cref{fig:model}. The modulated damping matrix becomes
\begin{gather}
     X =  (\eta_0 + \eta_1 + \eta_2) \cdot \mathbbm{1}_{n \times n} + i  \tilde{H}_\text{NH}. \label{eq:xth}
\end{gather}
Here, $\tilde{H}_\text{NH}$ is related to $H_\text{S}$ up to imaginary on-site potentials induced by sublattice dissipation:
\begin{gather}
    \tilde{H}_\text{NH} =  
  \begin{pmatrix}
     i |\eta_0|  & t_1 + \eta_1 & 0 & \\
    t_1 - \eta_1 & - i |\eta_0| & t_2 + \eta_2 & \\
    0 & t_2 - \eta_2 &  i |\eta_0| & \\
    & & & \ddots
  \end{pmatrix}.\label{eq:hnh}
\end{gather}
With a pure bath, we identify parameters in the distillation scenario:
\begin{gather}
    \eta_0 = \frac{1}{4} |\gamma^l_0|, \quad \eta_{1} = \frac{1}{2} |\gamma^l_1|, \quad \eta_{2} = \frac{1}{2} |\gamma^l_2|, \label{eq:setd}
\end{gather}
and in the amplification scenario:
\begin{gather}
    \eta_0 = -\frac{1}{4} |\gamma^g_0|, \quad \eta_{1} = -\frac{1}{2} |\gamma^g_1|, \quad \eta_{2} = -\frac{1}{2} |\gamma^g_2|. \label{eq:seta}
\end{gather} 

The advantage of a pure bath lies in the fact that one obtains a simple solution to  the covariance matrix $C_s$  in \Cref{eq:rss0}. It is easy to check that in the case of distillation ($M^g = 0$), 
\begin{gather}
     C_s = 0, \label{eq:cs0}
\end{gather}
and in the case of amplification ($M^g = M_b^g + M_0^g$),
 \begin{gather}
     C_s = -\mathbbm{1}.  \label{eq:cs1}
\end{gather}
For bosons, a hybrid bath with both bond loss and gain does not guarantee the existence of a unique covariance matrix, thus leading to singularities. We will resolve the solvable conditions for  a hybrid bath in \Cref{sec:is}.

\subsection{Exact Liouvillian spectra}
We proceed by  constructing the exact Liouvillian spectra for our model under OBC. From the mapping of \Cref{eq:xth}, we relate the rapidity spectrum $\beta_m$ of $X$ to the eigenvalues $\tilde{\epsilon}_m$ of $\tilde{H}_{\text{NH}}$: 
 \begin{gather}
 \beta_m = \eta_0 + \eta_1 + \eta_2 + i\tilde{\epsilon}_m.  \label{eq:beta}
 \end{gather}

The initial boundary mode of the NH SSH chain in \Cref{eq:bmm} remains an eigenmode of $\tilde{H}_{\text{NH}}$ in \Cref{eq:hnh} after \textcolor{black}{gap} modulation. 
One verifies that $ \tilde{H}_\text{NH} \  {\u{\psi}}_{R,0} = \tilde{\epsilon}_0 {{\u{\psi}}}_{R,0}$ at $\tilde{\epsilon}_0 = i|\eta_0|$.
The rapidity of the boundary mode becomes 
\begin{gather}
   \beta_0 = \eta_0 +\eta_1 + \eta_2  - |\eta_0|. \label{eq:beta0}
\end{gather}

For the bulk modes, their OBC spectrum can be obtained through a similarity transformation from the nonreciprocal $\tilde{H}_\text{NH}$ to a reciprocal Hamiltonian ${H}'_\text{NH}$ \cite{yao2018,yang2022,yang2024a}:
\begin{gather}
      {H}'_\text{NH} = S^{-1}  \tilde{H}_\text{NH} S = \begin{pmatrix}
     i |\eta_0|  & {t}'_1 & 0 & \\
    {t}'_1  & - i |\eta_0| & {t}'_2  & \\
    0 & {t}'_2  &  i |\eta_0| & \\
    & & & \ddots
  \end{pmatrix}, \label{eq:s0}
\end{gather}
where 
 \begin{gather} 
 S^T = 
     \begin{pmatrix}
         r & a r & r^2 & ar^2 & \cdots & r^{N}
     \end{pmatrix}. \label{eq:s}
 \end{gather}
Choosing rescaling parameters
\begin{gather}
   r = \sqrt{\frac{t_1^-t_2^-}{t_1^+t_2^+}}, \quad a =  \sqrt{\frac{t_1^-}{t_1^+}}, \label{eq:r}   
\end{gather}
one identifies
 \begin{gather}
     |t'_1| = |\sqrt{t_1^+t_1^-}|, \quad |t'_2| = |\sqrt{t_2^+t_2^-}|. \label{eq:t1p}
 \end{gather}
The reciprocal ${H}'_\text{NH}$ exhibits a gap closing between the boundary and bulk modes at
 \begin{gather}
    |t'_1| = |t'_2| \ \Leftrightarrow \  |t_1^2 - \eta_1^2| = |t_2^2 - \eta_2^2|. \label{eq:qpt}
 \end{gather}
Since eigenvalues do not change under the similarity transformation,  \Cref{eq:qpt} locates gapless phase transition lines of $\tilde{H}_{\text{NH}}$.  Comparing with \Cref{eq:pt} at $\eta_0 = 0$, it is noted that these transition lines are invariant under gap modulation.

Importantly, the nonunitary gauge transform encoded in the matrix $S$ of \Cref{eq:s} effectively modifies the Bloch phase factor: $e^{ikj} \to r^j e^{ikj}$. It allows to relate the OBC and PBC spectra of $\tilde{H}_{\text{NH}}$  through an imaginary momentum shift \cite{yang2024a}: 
\begin{gather}
    \tilde{\epsilon}^{\text{OBC}} (k) = \tilde{\epsilon}^{\text{PBC}} ( k - i\ln r ). \label{eq:ms}
\end{gather}
We thus validate \Cref{eq:gbz}, the result for  generic nonreciprocal Hamiltonians. At the same time, the gauge transform endows dynamical bulk skin modes with an exponential localization factor $r$ (\Cref{sec:smm}).

\begin{widetext}
Specifically, $\tilde{H}_{\text{NH}}$ respects  a spectral mirror symmetry 
$\tilde{\epsilon}^{\text{OBC}} (k) = \tilde{\epsilon}^{\text{OBC}} (-k)$, making the relation of \Cref{eq:ms} exact \cite{elisabet2020,yang2022}. 
We can build an exact GBZ with its non-Bloch Hamiltonian:  
 \begin{gather}
      \tilde{H}^{\text{non-Bloch}
    }_{\text{NH}}(k) =   \tilde{H}^{\text{Bloch}
    }_{\text{NH}}(k - i\ln r) 
    = \vec{d} \cdot \vec{\sigma}, 
\quad
    \begin{cases}
      & d_x = t_1 + t_2 \cos(k) + i\eta_2 \sin(k)  \\
     & d_y = i\eta_1 - i\eta_2 \cos(k) + t_2 \sin(k) \\
     & d_z = i|\eta_0|
    \end{cases}. \label{eq:nbh}
  \end{gather}
The eigen energies 
 share the form: $\tilde{\epsilon}_{\pm}(k) = \pm \sqrt{d_x^2 + d_y^2+d_z^2} = \pm d$,    
    \begin{gather}
     \tilde{\epsilon}_{\pm} (k) = \pm \sqrt{t_1^2+t_2^2-\eta_1^2-\eta_2^2 + \sgn{[(t_1+\eta_1)(t_2+\eta_2)]}2\sqrt{(t_1^2-\eta_1^2)(t_2^2-\eta_2^2)} \cos (k) - \eta_0^2}.  \label{eq:eps}
 \end{gather}
  \end{widetext}
A $\sgn$ function is added to produce the correct sign for our convention of $r$ in \Cref{eq:r}. Spectral mirror symmetry, manifested in $\cos (k)$,  enables one to 
take half of the BZ for discrete momenta: 
\begin{gather}
    k = \pi m/N \in (0,\pi), \quad m = 1,2, \dots, N-1. \label{eq:kv}
\end{gather} In this way, for a chain of $(2N-1)$ sites, apart from one boundary mode, we  retrieve the rapidity spectra of $2(N-1)$ bulk eigenmodes:
 \begin{gather}
 \beta_\pm (k) = \eta_0 + \eta_1 + \eta_2 + i\tilde{\epsilon}_\pm (k).  \label{eq:betab}
 \end{gather}
 It can be checked that the Liouvillian spectra in \Cref{eq:beta0} and \Cref{eq:betab}  are exact for arbitrary chain lengths $N \ge 2$.

\subsection{Liouvillian separation gap analysis}

\begin{figure}[t]
\centering
\includegraphics[width=0.95\columnwidth]{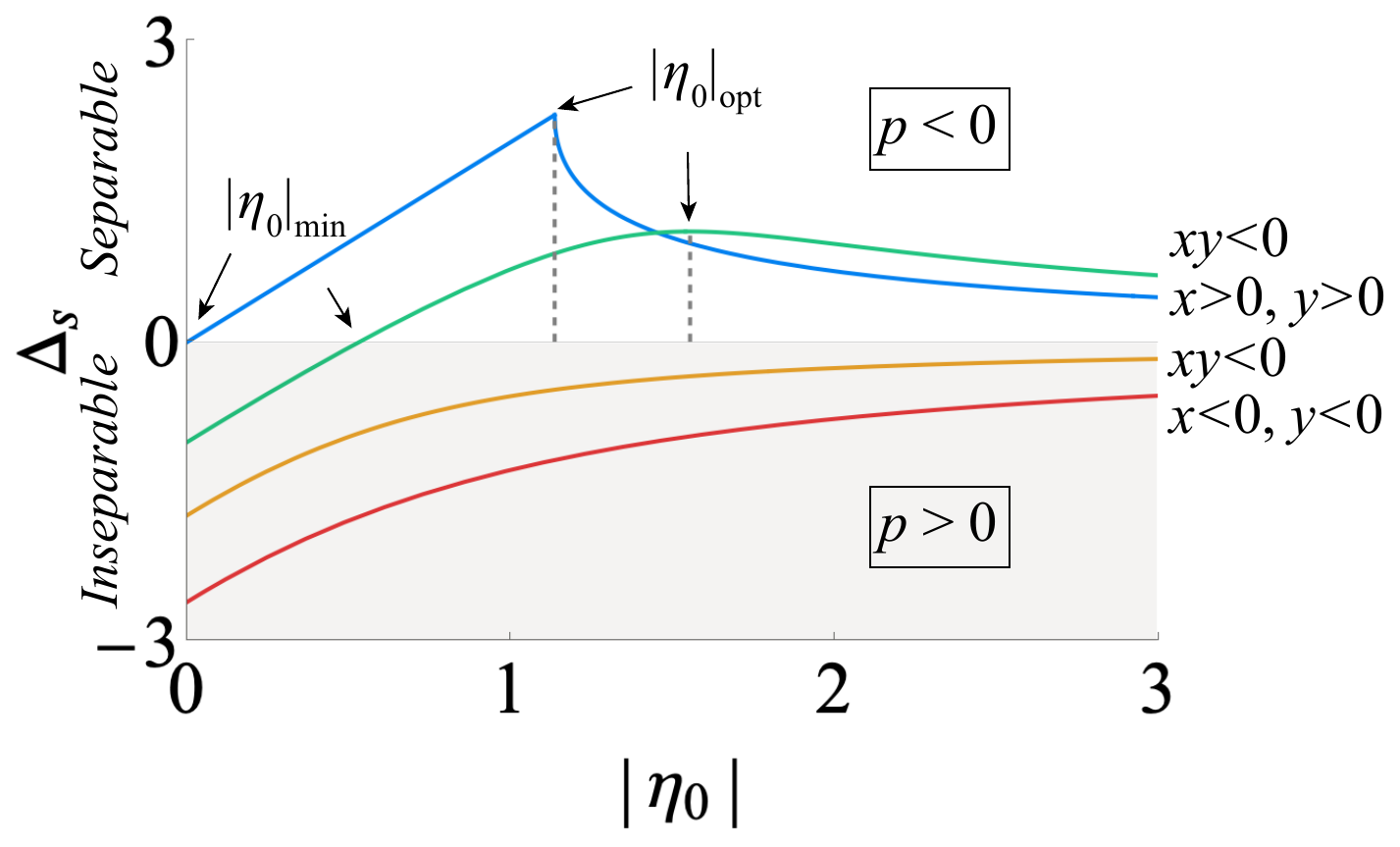}
\caption{{\bf Liouvillian separation gap as a function of sublattice dissipation $|\eta_0|$ in presence of weak ($p=\eta_1^2 + \eta_2^2 - t_1^2 - t_2^2 < 0$) and strong ($p > 0$) nonreciprocity.} We introduce auxiliary variables:  $x=|t_1|-|\eta_1|, y=|t_2|-|\eta_2|$.
$\Delta_s$ is evaluated from \Cref{eq:dbu}, the analytical solution in the large-$N$ limit.
We choose representative parameters for four curves with the notation $\vec{z} = (t_1,t_2,\eta_1,\eta_2)$: (blue) $\vec{z} = (1,2,0.5,0)$; (green) $\vec{z} = (0,1.5,0.5,0.2)$; (yellow) $\vec{z} = (0.5,0.5,1,0)$; (red) $\vec{z} = (0.25,0.5,0.5,1)$.}
\label{fig:sg}
\end{figure}

In this section, we will arrive at  a fully analytical criterion which predicts when a positive Liouvillian separation gap can arise in the NH SSH chain:
\begin{gather}
\Delta_s > 0, \quad  \text{if} \quad     \eta_1^2 + \eta_2^2 < t_1^2 + t_2^2, \quad    \textcolor{black}{|\eta_0| > |\eta_0|_{\min}}.
\label{eq:cri}
\end{gather}
Here, \textcolor{black}{a minimum sublattice dissipation strength $\eta_0$ is required when $(|t_1| - |\eta_1|)(|t_2| - |\eta_2|)<0$: $|\eta_0|_{\min} =\sqrt{(\eta_1^2-t_1^2)(t_2^2-\eta_2^2)/(t_1^2+t_2^2-\eta_1^2-\eta_2^2)}$. Otherwise,
$|\eta_0| >  0$ is sufficient.} \Cref{eq:cri} validates our general result in \Cref{eq:main} that the dynamical separability of NH boundary modes relies on both weak nonreciprocity and moderate engineered dissipation. In \Cref{fig:sg}, we further map out the development of $\Delta_s$  in full ranges of nonreciprocity and sublattice dissipation. In the separable region ($\eta_1^2 + \eta_2^2 < t_1^2 + t_2^2$), an optimal value of sublattice dissipation $|\eta_0|_\text{opt}$ exists which maximizes  $\Delta_s$. 

\begin{figure}[t]
\centering
\includegraphics[width=0.85\columnwidth]{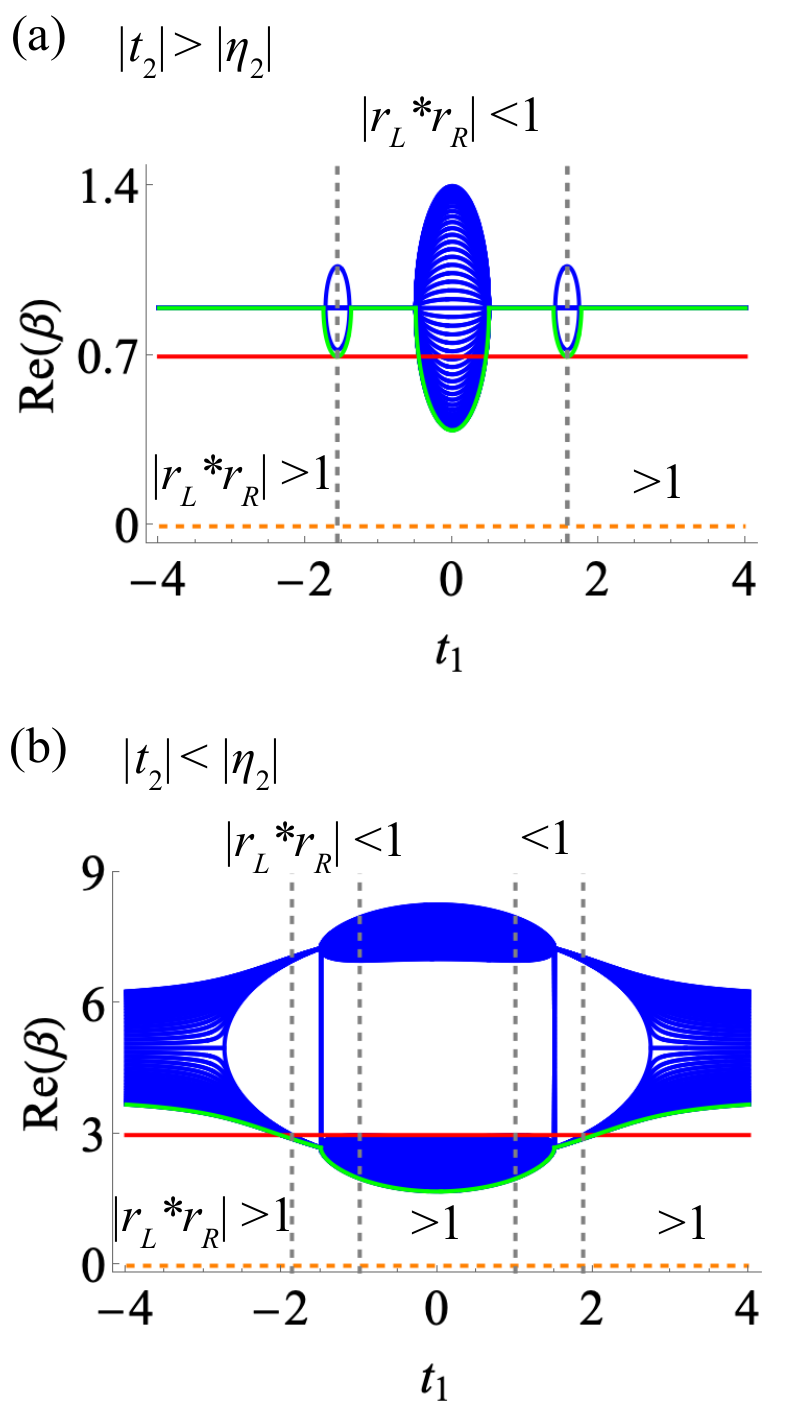}
\caption{{\bf Bulk Liouvillian gap developed from the bottom of the real rapidity spectrum.} The green lines denote one half of bulk Liouvillian gap ($\Delta_{\text{bulk}}/2$)  in the large-$N$ limit, analytically resolved  in \Cref{eq:dbu} by the relation $\Delta_\text{bulk} = \Delta_s + \Delta_{\text{boundary}}$. The blue lines show the real rapidity spectrum of bulk modes from exact solutions in \Cref{eq:betab} and \Cref{eq:eps} for a finite odd-length chain $n_\text{tot} = 2N-1$ with $N=46$. The NH boundary mode is depicted by the red line with a rapidity given in \Cref{eq:beta0}.
We vary $t_1$ and fix $t_2 =1.5, \eta_1 = 0.5, \eta_2 = 0.2$, $\eta_0 = 0.2$ in (a) and $t_2 =1, \eta_1 = \eta_2 = 1.5$, $\eta_0 = 2$ in (b). The phase transitions occur at $|r_L^*r_R| = 1$, corresponding to $t_{1,c} = \pm \sqrt{t_2^2-\eta_2^2+\eta_1^2} \simeq \pm 1.568$ in (a) and $t_{1,c} = \pm \sqrt{\pm (t_2^2-\eta_2^2)+\eta_1^2} \simeq \pm 1, \pm 1.871$ in (b).}
\label{fig:gap}
\end{figure}

Our analysis is based on the large-$N$ limit where the Liouvillian separation gap can be analytically resolved.  For a finite system size, $\Delta_s$  converges to this limit quickly (\Cref{fig:gap}).
By definition [\Cref{eq:sgap} and \Cref{eq:lgbb}],  $\Delta_s$ measures the distance between real rapidity spectra of bulk and boundary modes [\Cref{eq:betab} and \Cref{eq:beta0}]:
\begin{gather}
    \Delta_s =2 \min_{k, \pm}  {\re [|\eta_0| + i\tilde{\epsilon}_\pm(k)]}.
\end{gather}
It is easy to discern that the extrema of bulk energies $\tilde{\epsilon}_\pm(k)$ in \Cref{eq:eps}  are  reached at $\cos(k)= \pm 1$, or  $k = 0, \pi$. When $N$ is large, 
the discrete momenta in the exact solutions of \Cref{eq:kv} can approach these values. We thus make the estimation:
\begin{gather}
   \Delta_s \overset{N \to \infty}{\simeq} 2 |\eta_0| + 2\min_{\pm} \{ \re [ i\tilde{\epsilon}_\pm(0)], \re [ i\tilde{\epsilon}_\pm(\pi)]\}.
\end{gather}
It enables us to obtain an analytical solution to the Liouvillian separation gap.

First, we look at the case without sublattice dissipation ($\eta_0 = 0$): 
  \begin{gather}
     \Delta_s =  - 2\sum_{i=1,2} \theta(|\eta_i| - |t_i|)\sqrt{\eta_i^2 - t_i^2}, \label{eq:d0}
  \end{gather}
where $\theta (x)$ denotes the Heaviside step function: $\theta (x) = 0$ for $x < 0$ and $\theta (x) = 1$ for  $x > 0$.
Given arbitrary $t_i, \eta_i \in \mathbb{R}$, $\Delta_s \le 0$ consistent with our prediction from $P$ symmetry [\Cref{eq:chiral} and \Cref{eq:hch}].  $\Delta_s = 0$ is reached when $|t_1| > |\eta_1|$ and $|t_2| > |\eta_2|$, as shown by the vertical axis in \Cref{fig:sg}. In any case,  the NH boundary mode cannot become the single longest-lived mode without gap modulation.

Including the sublattice dissipation ($\eta_0 \ne 0$), we find analytical expressions of Liouvillian separation gap in three distinct regions: \begin{widetext}
  \begin{align}
  \text{I.} \ &|t_1| > |\eta_1| \ \text{and} \  |t_2| > |\eta_2|, \notag \\
     &\Delta_s = 
          2|\eta_0|  - 2\theta{(|\eta_0| - |\sqrt{t_1^2 - \eta_1^2}-\sqrt{t_2^2 - \eta_2^2} |)}\sqrt{\eta_0^2 - (\sqrt{t_1^2 - \eta_1^2}-\sqrt{t_2^2 - \eta_2^2})^2}; \notag \\
     \text{II.} \ &|t_1| < |\eta_1| \ \text{and} \  |t_2| > |\eta_2| \ \  \text{or} \ \  |t_1| > |\eta_1| \ \text{and} \  |t_2| < |\eta_2|, \notag \\
     &\Delta_s = 
           2|\eta_0| - \sqrt{2(\eta_0^2 +\eta_1^2 + \eta_2^2 - t_1^2 -t_2^2+ \sqrt{\eta_0^4+2\eta_0^2 (\eta_1^2 + \eta_2^2 - t_1^2 -t_2^2)+(\eta_1^2 -t_1^2 +t_2^2 - \eta_2^2)^2})}; \notag  \\
             \text{III.} \ &|t_1| < |\eta_1| \ \text{and} \  |t_2| < |\eta_2|, \notag \\
     &\Delta_s = 
           2|\eta_0| - 2\sqrt{\eta_0^2 + (\sqrt{\eta_1^2 - t_1^2}+\sqrt{\eta_2^2 - t_2^2})^2}.   \label{eq:dbu} 
  \end{align}
  \end{widetext}
Taking the limit $\eta_0 \to 0$ in each region, one recovers our earlier result in \Cref{eq:d0}. It is useful to check the validity of \Cref{eq:dbu} in all three regions by comparing it with exact solutions calculated at a finite system size. 
In \Cref{fig:gap}, we locate the minimum of bulk real rapidity spectrum (green) by measuring half of bulk Liouvillian gap: $\Delta_\text{bulk}/2 = (\Delta_s + \Delta_{\text{boundary}})/2 = \Delta_s/2 + \beta_0$, directly from \Cref{eq:dbu}. Our analytical solution  for large system size $N$  (green) agrees well with finite-size results at $N=46$ (blue).

Starting from \Cref{eq:dbu}, we are ready to examine in each parameter region  whether a positive $\Delta_s$ can be achieved by tuning sublattice dissipation $\eta_0$. If successful, NH boundary mode will be dynamically selected during relaxations.  Critical points in \Cref{eq:pt} require special treatment, at which bulk and boundary eigenmodes become degenerate. For simplicity, we choose to first exclude them in the general analysis of $\Delta_s (\eta_0)$. Later, we recover the asymptotic behavior of $\Delta_s$ when approaching these critical points.

\subsubsection{Region I: \  $|t_1| > |\eta_1|, |t_2| > |\eta_2|$}
In the first region of \Cref{eq:dbu},
when $|t_1| > |\eta_1|$ and $|t_2| > |\eta_2|$, one verifies
\begin{gather}
\Delta_s > 0, \quad \text{if} \ |\eta_0| > 0. \label{eq:re1}
\end{gather}
It is an ideal case where an infinitesimal $\eta_0$ helps select the NH boundary mode effectively (\Cref{fig:sg}, blue curve). As $|\eta_0|$ slowly increases from zero, $\Delta_s = 2|\eta_0|$ keeps a linear growth, ensuring that the NH boundary mode becomes more predominant in long-time relaxation dynamics. Once $|\eta_0| > |\sqrt{t_1^2 - \eta_1^2}-\sqrt{t_2^2 - \eta_2^2} |$, we learn from its derivative 
\begin{gather}
    \partial_{|\eta_0|} \Delta_s  = 2- \frac{2|\eta_0|}{\sqrt{\eta_0^2 - (\sqrt{t_1^2 - \eta_1^2}-\sqrt{t_2^2 - \eta_2^2})^2}} < 0,
\end{gather}
that $\Delta_s$ starts to decrease.
Therefore, we obtain an optimal $\eta_0$ which maximizes $\Delta_s$:
 \begin{gather}
     \Delta_{s,\text{max}} = 2|\sqrt{t_1^2 - \eta_1^2}-\sqrt{t_2^2 - \eta_2^2}| = 2 |\eta_0|_\text{opt}. \label{eq:opt1}
 \end{gather}
Both in the large-$N$ limit (\Cref{fig:sg}, blue curve) and at a finite $N$ [\Cref{fig:model}~(a) and (b)], the optimal sublattice dissipation strength  ($|\eta_0|_\text{opt} \simeq 1.134$) produces a maximal Liouvillian separation gap ($     \Delta_{s,\text{max}}  \simeq 2.268$).

\subsubsection{Region II: \ $|t_1| < |\eta_1|, |t_2| > |\eta_2|$ or $|t_1| > |\eta_1|, |t_2| < |\eta_2|$}

Slightly increasing nonreciprocity by allowing $|t_1| < |\eta_1|$ or $|t_2| < |\eta_2|$, we enter the second region of \Cref{eq:dbu}, where the main criterion on the dynamical separability given by \Cref{eq:cri} can be drawn. 

It is convenient to introduce two auxiliary functions:
\begin{gather}
p = \eta_1^2 + \eta_2^2 - t_1^2 -t_2^2, \quad q=\eta_1^2 - t_1^2 + t_2^2  - \eta_2^2, 
\end{gather}
which satisfy $q^2 - p^2 = -4 (\eta_1^2 - t_1^2)(\eta_2^2 - t_2^2) > 0$.
Liouvillian separation gap is simplified to
\begin{gather}
    \Delta_s = 2|\eta_0| - \sqrt{2(\eta_0^2+p+\sqrt{\eta_0^4+2p\eta_0^2+q^2})}. \label{eq:deta0}
\end{gather}
One immediately identifies an inseparable region:
\begin{gather}
   \Delta_s < 0, \quad \text{if} \  p > 0, \ \forall 
\eta_0, \label{eq:re2}
\end{gather}
consistent with our criterion in \Cref{eq:cri}.
As can be seen from \Cref{fig:sg} (yellow curve), in this region, regardless of the strength of sublattice dissipation $|\eta_0|$, the negative sign of $\Delta_s$ cannot be flipped. NH bulk modes always prevail over the boundary mode during relaxations. 

Whereas, when $p < 0$, Liouvillian separation gap can be closed by a nonzero $\eta_0$:
\begin{gather}
     \Delta_s = 0, \quad \text{at} \  |\eta_0|_{\text{min}} = \sqrt{(q^2-p^2)/(-4p)}, \label{eq:eta0min}
\end{gather} 
as shown in \Cref{fig:sg} (green curve). One can look in more detail at the derivative:
\begin{gather}
    \partial_{|\eta_0|} \Delta_s 
    = 2 -\frac{2 |\eta_0|}{(2|\eta_0|-\Delta_s)}(1+\frac{\eta_0^2 + p}{\sqrt{\eta_0^4+2p\eta_0^2+q^2}}).
\end{gather}
For $0 < |\eta_0| < |\eta_0|_{\text{min}}$, we have
 \begin{gather}
   \partial_{|\eta_0|} \Delta_s 
  > 1-\frac{\eta_0^2+p}{\eta_0^2-p} > 0,
\end{gather}
valid when $p<0$. It shows that before reaching zero at the gap closing point, 
$\Delta_s$ increases monotonically with $|\eta_0|$.
On the other hand, for $|\eta_0| > |\eta_0|_{\text{min}}$, $\partial_{|\eta_0|}\Delta_s$ holds an upper non-negative bound:
\begin{gather}
   \partial_{|\eta_0|} \Delta_s < 1- \frac{\eta_0^2+p}{\eta_0^2-p}. \label{eq:ddelta}
\end{gather}
Taking the limit $|\eta_0| \to \infty$ in \Cref{eq:ddelta} and \Cref{eq:deta0}, one recognizes
\begin{gather}
 \lim_{|\eta_0| \to \infty} \partial_{|\eta_0|} \Delta_s < 0, \quad  \lim_{|\eta_0| \to \infty} \Delta_s = 0.
\end{gather}
By increasing $|\eta_0|$, $\Delta_s$ first reaches the maximal positive value, then drops. Also, when $|\eta_0| > |\eta_0|_{\text{min}}$, $\Delta_s$ stays positive due to the lower bound at $|\eta_0| \to \infty$  (\Cref{fig:sg}, green curve). 
One thus arrives at the condition for a positive $\Delta_s$ in the second region: 
\begin{gather}
  \Delta_s > 0, \quad \text{if} \  p < 0, \  |\eta_0|_\text{min}  < |\eta_0| \ll \infty, \label{eq:emin}
\end{gather}
where $|\eta_0|_\text{min}$ is given by \Cref{eq:eta0min}.

We can also determine the 
 optimal $\eta_0$ from a vanishing derivative. A straightforward  calculation leads to
 \begin{gather}
     \partial_{|\eta_0|} \Delta_s = 0 \quad \Leftrightarrow \quad a\eta_0^6 +b\eta_0^4+c\eta_0^2+d=0,
 \end{gather}
with $a=4p, b= 9p^2+3q^2, c = 12pq^2$ and $d=4q^4$.
Our task is simplified to finding the real roots of the cubic equation of $\eta_0^2$. 
It turns out that 
 \begin{gather}
   (\Delta_s)_\text{max}  = \Delta_s(|\eta_0|_{\text{opt}}), \label{eq:cubic} \\  |\eta_0|_{\text{opt}} = \left[\frac{1}{4p}(h(p,q) - \frac{g(p,q)}{9h(p,q)}-3p^2-q^2)\right]^{1/2}, \notag
 \end{gather}
where we introduce 
$g$ and $h$ functions: $g(p,h) = \textcolor{black}{-81p^4}+90p^2q^2-9q^4$ 
and $h(g,h) = ( -27p^6 +45p^4q^2 -17p^2q^4-q^6+8\sqrt{p^6q^6 - 2p^4q^8+p^2q^{10}})^{1/3}$. For the chosen parameters in \Cref{fig:sg} (green curve), we identify through \Cref{eq:cubic}: $ |\eta_0|_{\text{opt}} \simeq 1.552$, $   (\Delta_s)_\text{max} \simeq 1.103$.

\subsubsection{Region III: \  $|t_1| < |\eta_1|, |t_2| < |\eta_2|$}

In the third region of \Cref{eq:dbu}, 
when $|t_1| < |\eta_1| \ \text{and} \  |t_2| < |\eta_2|$, it is clear that
\begin{gather}
    \Delta_s < 0, \quad \forall \eta_0. \label{eq:re3}
\end{gather}
It is no longer possible to apply sublattice dissipation to elevate the NH boundary mode (\Cref{fig:sg}, red curve). Intuitively, as can be seen in \Cref{fig:gap}~(b), in this region ($|t_1| < \eta_1 = 1.5$), strong nonreciprocity from bond dissipation embeds the boundary mode (red) too deep inside the real rapidity spectrum of bulk modes (blue).

Combining results in three regions [\Cref{eq:re1}, \Cref{eq:re2}, \Cref{eq:emin}  and \Cref{eq:re3}] , we arrive at \Cref{eq:cri}, the analytical criterion for a positive Liouvillian separation gap in the NH SSH chain.

In the end, we reveal the asymptotic behavior of $\Delta_s$ in the separable region ($\eta_1^2 + \eta_2^2 < t_1^2 + t_2^2$) when the system 
approaches gapless phase transition points ($|t_1^2 - \eta_1^2| = |t_2^2 - \eta_2^2|$). It is sufficient to study the tendency of special points in the upper plane of \Cref{fig:sg}. 
For $|t_1| >  |\eta_1|$, $|t_2| >  |\eta_2|$, \textcolor{black}{we look at the maximal value of $\Delta_s$.
From \Cref{eq:opt1}, $\Delta_{s,\text{max}} = 0$ at $t_1^2 - \eta_1^2 = t_2^2 - \eta_2^2$. It indicates that towards phase transitions, the blue curve tends to collapse to a flat zero line: $\Delta_s = 0$,  $\forall \eta_0$}. 
It is consistent with a vanishing Liouvillian separation gap in \Cref{fig:gap}~(a) at $t_{1,c} \simeq \pm 1.568$. For  $(|t_1| - |\eta_1|)(|t_2| - |\eta_2|) < 0$, critical points occurring at $p =\eta_1^2 + \eta_2^2 - t_1^2 - t_2^2 = 0$ push the green curve in \Cref{fig:sg}  outside of the separable region. Indeed,
from \Cref{eq:eta0min}, $|\eta_0|_{\text{min}} \to \infty$ as $p \to 0$. The green curve remains below the zero line: $\Delta_s < 0$, $\forall \eta_0$. It agrees with a negative Liouvillian separation gap in \Cref{fig:gap}~(b) at $t_{1,c} \simeq \pm 1.871$. 

From the analysis above, if we aim to observe the BBC through the measurement of dynamical NH boundary mode, apart from \textcolor{black}{satisfying the criterion for a positive Liouvillian separation gap in \Cref{eq:cri},
 gapless phase transitions should be confined to the region:} 
\begin{gather}
t_1^2 - \eta_1^2  = t_2^2 - \eta_2^2. \label{eq:ps1}
\end{gather} 
In this way,  $\Delta_s$ vanishes exactly at these critical points while stays positive everywhere else. Any dynamical observable capturing $\Delta_s = 0$ (e.g. polarization drift in \Cref{sec:pol})  can then be used as a witness of the NH BBC [\Cref{eq:pt}].

\section{Dynamical signatures}
\label{sec:dyn}
Next, we embark on identifying dynamical signatures of the NH boundary mode. 
The time evolution of boson density is constructed from exact skin eigenmodes of the damping matrix. It enables us to observe distinctive chiral damping phenomena  before and after the separation of NH bulk and boundary modes, respectively. On top of that, we propose a key observable, polarization drift, as a dynamical witness of the biorthogonal BBC in the NH SSH chain.

\subsection{Skin modes under gap modulation}
\label{sec:smm}
It is convenient to extract the boson density from the diagonal part of the correlation matrix in \Cref{eq:tc}: $n_l(t) = \tr[\rho(t) b_l^\dagger b_l] = C_{ll}(t)$, where $l$ denotes the $l$-th site. For our setups [\Cref{eq:xth}-\Cref{eq:seta}], one can choose the initial state as a uniformly filled chain with $x$ bosons per site ($x \ge 0$): $C(0) = x \times \mathbbm{1}$. In the distillation ($\eta_0 > 0$, $C_s = 0$) and  amplification  ($\eta_0 < 0$, $C_s = -\mathbbm{1}$) scenarios, the boson density follows the time evolution 
\begin{gather}
  n_l(t) = \begin{cases}
       x  Q(l,t), & \eta_0 > 0 \\
     -1 +  (x+1)  Q(l,t), &   \eta_0 < 0
  \end{cases}  \label{eq:den} 
\end{gather}
with the propagator 
\begin{align}
  Q(l,t) = 
&\sum_{m,m'} \sum_{l'=1}^{n_{\text{tot}}}  e^{-(\beta_m + \beta^*_{m'} )t} \notag \\
& \times \u{\tilde{\psi}}_{Lm'}(l)\u{\tilde{\psi}}^*_{Rm'}(l')\u{\tilde{\psi}}_{Rm}(l')\u{\tilde{\psi}}^*_{Lm}(l).  \label{eq:prop}
\end{align}
Band indices $m$ and $m'$ sum over all of skin eigenmodes $\tilde{\psi}_{R/L}$ of the damping matrix under gap modulation [\Cref{eq:xth} and \Cref{eq:hnh}]. The 
damping matrix $X$ shares the same eigenmodes with $\tilde{H}_\text{NH}$. 

For the boundary mode ($m=0$) in \Cref{eq:bmm}, we have verified that it remains an eigenmode before and after the modulation with a rapidity $\beta_0$ given by \Cref{eq:beta0}.

As to bulk modes [$m = (\pm,k )$] endowed with rapidities  $\beta_\pm (k)$ in  \Cref{eq:betab}, we will see  in presence of sublattice dissipation, the skin-effect localization profile keeps the same but the non-Bloch components change.  

On one hand, the localization tendency of  bulk modes
can be read directly from  the similarity transformation [\Cref{eq:s0}-\Cref{eq:t1p}]:
\begin{gather}
    \tilde{\u{\psi}}_{R,m} = S\u{\psi}'_{R,m}, \quad \tilde{\u{\psi}}_{L,m} = (S^*)^{-1}\u{\psi}'_{L,m}. 
\end{gather}
While $\u{\psi}'_{R/L}$ are 
normal bulk eigenmodes of the reciprocal Hamiltonian $H'_\text{NH}$ devoid of the NHSE, the nonunitary gauge transform $S$ results in
\begin{align}
     \tilde{\u{{\psi}}}_{R,m} (j,A/B) &\propto  r^j, \notag \\
      \tilde{\u{{\psi}}}_{L,m} (j,A/B) &\propto  (r^*)^{-j}, \label{eq:seb}
\end{align} 
where $(j, \alpha)$ stands for the site $l=2j-1 $ ($l = 2j$) on the sublattice $\alpha = A \ (B)$ of the $j$-th unit cell. Indeed, for bulk modes, the localization  parameter given by $r$ in \Cref{eq:r} is 
independent of sublattice dissipation ($\eta_0$).

On the other hand, the non-Bloch components of bulk modes are determined by biorthogonal eigenvectors of the non-Bloch Hamiltonian in \Cref{eq:nbh}:
\begin{align}
     \u{u}^{\text{OBC}}_{R,\pm}(k) &= \frac{1}{\sqrt{2d(d\mp d_z)}} \begin{pmatrix}
         d_x - id_y \\ \pm d - d_z
     \end{pmatrix}, \notag \\
     \u{u}^{*\text{OBC}}_{L,\pm}(k) &= \frac{1}{\sqrt{2d(d\mp d_z)}} \begin{pmatrix}
         d_x + id_y \\ \pm d - d_z 
     \end{pmatrix}, \label{eq:url}
 \end{align}
satisfying $\u{u}_{L,\alpha}^*(k) \cdot \u{u}_{R, 
\alpha'}(k) = \delta_{\alpha, \alpha'}$. Sublattice dissipation contributes to a nonzero component $d_z = i|\eta_0|$.  

Combined with analyses in \Cref{eq:seb} and \Cref{eq:url}, we are ready to construct exact  bulk eigenmodes of the damping matrix. It should be noted that another constraint comes from the
spectral mirror symmetry: $\beta_{\pm}(k) = \beta_{\pm}(-k)$. This double degeneracy implies a superposition of non-Bloch waves at opposite momenta:
\begin{gather}
  \tilde{\u{{\psi}}}_{R,(\pm,k)} (j)
  =  \frac{ r^j }{\sqrt{2N}} [ e^{ikj} \u{u}^{\text{OBC}}_{R,\pm} (k) - e^{-ikj} \u{u}^{\text{OBC}}_{R,\pm} (-k) ], \notag \\
   \tilde{\u{{\psi}}}_{L,(\pm,k)} (j) =   \frac{(r^*)^{-j}}{\sqrt{2N}}  
[ e^{ikj} \u{u}^{\text{OBC}}_{L,\pm} (k) - e^{-ikj} \u{u}^{\text{OBC}}_{L,\pm} (-k) ]. \label{eq:psib}
\end{gather}
In our notation, for a site in the $j$-th unit cell, individual weights on $A$ and $B$ sublattices are encoded in $\u{u}_{R/L}$ of \Cref{eq:url}.
The relative minus sign between the two non-Bloch waves meets the open boundary condition: the last  unit cell is broken with an empty $B$ site, 
\begin{gather}
   \tilde{\u{{\psi}}}_{R,(\pm,k)} (N,B) =  \tilde{\u{{\psi}}}_{L,(\pm,k)} (N, B) = 0, \quad \forall k. 
\end{gather} 

It can be checked that the boundary and bulk eigenmodes constructed in \Cref{eq:bmm} and \Cref{eq:psib} mutually respect biorthogonal relations:
\begin{gather}
    \u{\psi}^*_{L,m}  \cdot \u{\psi}_{R,m'}  = \delta_{m,m'}.
\end{gather} 
Also, our exact eigenmodes are valid for arbitrary real parameters $t_i^\pm, \gamma_i^\pm \in \mathbb{R}$ and  arbitrary system sizes   $n_\text{tot} = 2N-1$ ($N \ge 2$).

\subsection{Selective damping in boson density}
\begin{figure}[t]
\centering
\includegraphics[width=1\columnwidth]{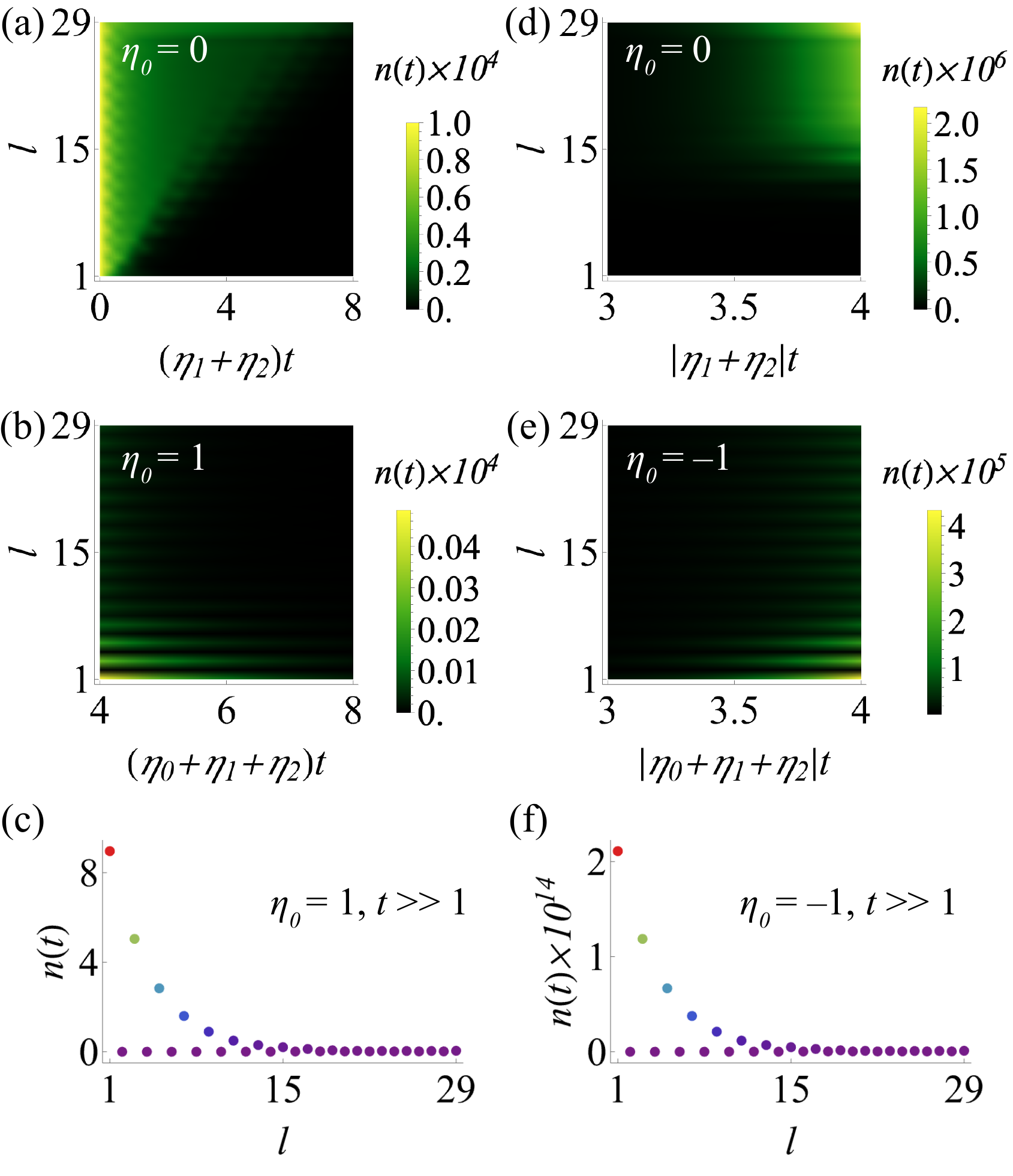} 
\caption{{\bf Bulk and boundary Liouvillian skin effects dynamically selected by sublattice dissipation.} 
In the distillation (a)–(c) and amplification (d)–(f) scenarios, we take the same set of hopping parameters $\vec{z} = (t_1, t_2, \eta_1, \eta_2)$  as in \Cref{fig:model}~(a) and (b), and then compare the results with and without sublattice dissipation: $\eta_0 = 0, \pm 1$. The initial state  is chosen for (a)-(c) at $x = 10^4$ bosons per site  and for (d)-(f) at $x=0$ (an empty chain). Sites with odd (even) index $l$ belong to the sublattice $A$ ($B$). At long times, boson density profile along the chain highlights different localization tendencies of NH bulk modes with $|r|^{-1} \simeq 1.73 > 1$ in (a) and (d) and NH boundary mode with  $|r_L| = 0.75 < 1$ in (b) and (e). 
 In (c) and (f), 
with $|\eta_0|=1$ close to $|\eta_0|_{\text{opt}} = 1.134$ which maximizes a positive $\Delta_s$ (\Cref{fig:sg}, blue curve), the long-time density at $|\eta_0+\eta_1+\eta_2|t = 10$ is dominated by the NH boundary mode [\Cref{eq:bmm}].
}
\label{fig:bm}
\end{figure}
From \Cref{eq:den} and \Cref{eq:prop}, it follows that at long times   the localization profile of boson density along the chain is determined by the left eigenmode $\u{\tilde{\psi}}_{L,\tilde{m}}$ with the smallest real rapidity: $\re[{\beta_{\tilde{m}}}] \le \re[{\beta_{m}}], \forall m \ne\tilde{m}$. 
Suppose  $\u{\tilde{\psi}}_{L,\tilde{m}}(l) \propto  \zeta_{L,\tilde{m}}^{l}$, it leads to
\begin{gather}
|n_{l}(t)| \propto |\zeta_{L,\tilde{m}}|^{2l} e^{-\Delta t}, \label{eq:nzeta}
\end{gather}
of which in the time-dependent part  we retrieve the Liouvillian gap:
\begin{gather}
    \Delta = 2\re [{\beta_{\tilde{m}}}].
\end{gather}

\subsubsection{Bulk Liouvillian skin effects: $\Delta_s \le 0$}
 In absence of sublattice dissipation ($\eta_0 = 0$), Liouvillian separation gap in \Cref{eq:d0} remains non-positive. It is  NH bulk modes [\Cref{eq:psib}] that share smaller real rapidities and  contribute most to the boson density: 
\begin{gather}
|n_{j,A/B}(t)| \propto |r|^{-2j} e^{-\Delta_{\text{bulk}} t}. \label{eq:ntb}
\end{gather}
From \Cref{eq:lgbb} and \Cref{eq:betab},  the bulk Liouvillian gap reads
\begin{gather}
    \Delta_{\text{bulk}} = 2(\eta_1 + \eta_2) - 2\sum_{i=1,2} \theta(|\eta_i| - |t_i|)\sqrt{\eta_i^2 - t_i^2}. \label{eq:dbulk}
\end{gather}

In the case of distillation in \Cref{eq:setd}, $\eta_1, \eta_2 > 0$, one discerns $ \Delta_{\text{bulk}} > 0$. The stability of the system is ensured by bond loss.  
Taking the limit $t \to \infty$ in \Cref{eq:ntb}, the system relaxes to the steady state of an empty chain: $n_{s,l} = C_{s,ll} = 0$. When $|r|^{-1} > 1$, the particle at the right end of the chain exhibits the longest lifetime [\Cref{fig:bm}~(a)]. One can estimate its lifetime by $|n_{N,A}(\tau)| = e^{-1} \ll 1$. It turns out that
\begin{gather}
    \tau \propto \frac{1}{ \Delta_{\text{bulk}} } + \frac{1}{\xi} \cdot \frac{(2N-1)}{ \Delta_{\text{bulk}} }, \label{eq:tau}
\end{gather}
where $\xi > 0$ denotes the correlation length of the system: 
\begin{gather}
    \xi = |\ln |r||^{-1}.
\end{gather}
The total lifetime of boson  grows linearly with the system length without closing the bulk Liouvillian gap. We thus observe the bulk Liouvillian skin effect \cite{ueda2021,yang2022}. 
In addition, the lifetime of boson on each individual site can also be estimated from \Cref{eq:tau} by replacing the total length $(2N-1)$ with the site index $l$. The time lapse for the particle wave to terminate increases linearly with $l$, leading to a chiral damping wavefront \cite{fei2019,yang2022} in \Cref{fig:bm}~(a).

In the case of amplification in \Cref{eq:seta}, $\eta_1, \eta_2 < 0$, $ \Delta_{\text{bulk}} < 0$. The system becomes unstable due to bond gain. When $t \to \infty$, the boson density in \Cref{eq:ntb} diverges without reaching the reference steady state: $n_{s,l} = C_{s,ll} = -1$. Nevertheless, resulting from the prefactor $|r|^{-2j}$, the bulk Liouvillian skin effect  still manifests itself in the unstable region such that at a given time,  one end of the chain is populated by an exponentially growing number of bosons compared to the other end of the chain [\Cref{fig:bm}~(d)]. 

\subsubsection{Boundary Liouvillian skin effects: $\Delta_s > 0$}
In presence of sublattice dissipation ($\eta_0 \ne 0$), guided by the criterion in \Cref{eq:cri},  we are able to engineer a positive Liouvillian separation gap in the separable region of weak nonreciprocity ($\eta_1^2 +\eta_2^2 < t_1^2 + t_2^2$).

Once $\Delta_s > 0$, the long-time dynamics ($t \gg 1$) is governed by 
 NH boundary mode [\Cref{eq:bmm}], resulting in
\begin{align}
&n_{j,A}(t) \propto |r_L|^{2j} e^{-\Delta_{\text{boundary}} t}, \notag \\
&n_{j,B}(t)/n_{j,A}(t) \simeq 0. \label{eq:nts}
\end{align}
From \Cref{eq:lgbb} and \Cref{eq:beta0}, the boundary Liouvillian gap is given by 
\begin{gather}
    \Delta_{\text{boundary}} =2 (\eta_1 + \eta_2+\eta_0 - |\eta_0|).
\end{gather}

In the distillation scenario, $\eta_{i=0,1,2} >0$, the system is stable with $\Delta_{\text{boundary}} > 0$.
Before it finally relaxes to an empty chain, 
the system is able to acquire the density profile of NH boundary mode, which is exponentially localized on the sublattice $A$ and fully suppressed on the sublattice $B$: $n_{j,B}(t) = 0$. Setting $\Delta_s > 0$ by tuning sublattice dissipation $\eta_0 = 1$ [\Cref{fig:model}~(a)],  in \Cref{fig:bm}~(b) and (c), we observe features characteristic of the boundary Liouvillian skin effect. Compared with its bulk counterpart at $\eta_0 = 0$ [\Cref{fig:bm}~(a)], the localization direction of boson density is now switched from the right ($|r|^{-1} > 1$) to the left ($|r_L| < 1$)  end of the chain. 

In the amplification scenario, one gets $\eta_{i=0,1,2} < 0$ and $\Delta_{\text{boundary}} < 0$. In this case, the system becomes unstable with diverging boson densities on both sublattices. Still, beyond short times, NH boundary mode becomes the fastest growing mode ($\Delta_s > 0$): $\re[\beta_0] < \re[\beta_m] < 0, \forall m \ne 0$, see  $\eta_0 = -1$ in \Cref{fig:model}~(b). The density profiles in \Cref{fig:bm}~(e) and (f) then retain main features of NH boundary mode with comparatively negligible distribution on the sublattice $B$: $n_{j,B}(t)/n_{j,A}(t) \simeq 0$.  
 Apart from the switching of the localization direction, notably,  the boundary Liouvillian skin effect yields a highly populated and sharply localized NH boundary mode at long times [\Cref{fig:bm}~(f)]. It suggests potential application in NHSE-enhanced single-mode lasers~\cite{zhu2022}.\\

%
\subsection{Polarization drift and  NH BBC}
\label{sec:pol}

With access to NH boundary mode, 
our separation scheme makes it possible to directly observe the NH BBC in open quantum systems. 

As a prerequisite,  we recall the functioning parameter space for $\{t_i, \eta_i \}$ [\Cref{eq:cri} and \Cref{eq:ps1}]. It requires weak nonreciprocity and  moderate sublattice dissipation: $\eta_1^2 +\eta_2^2 < t_1^2 + t_2^2$ and $|\eta_0| > |\eta_0|_\text{min}$, so that in gapped phases we have $\Delta_s > 0$.  At the same time, the gapless phase transition lines should be confined to $t_1^2 - \eta_1^2  = t_2^2 - \eta_2^2$, along which $\Delta_s = 0$ faithfully captures the degeneracy between NH bulk and boundary modes. 

In closed systems (\Cref{sec:sym}), given a pure NH boundary mode, the BBC is manifested in an integer topological invariant, the biorthogonal polarization $\mathcal{P}_0$ [\Cref{eq:pol}].
It is convenient to extend the notion to open systems. We introduce the dynamical polarization of a mixed state from the density map of the system~\cite{yang2022,meng2025}:
\begin{gather}
      \mathcal{P} (t) = \frac{\sum_{l=1}^{n_{\text{tot}}} l \cdot {n}_l(t)}{{n_{\text{tot}}} \cdot \sum_{l=1}^{n_{\text{tot}}} {n}_l(t)}. \label{eq:dpol}
\end{gather}
With this definition, $\mathcal{P} (t) \in [0, 1]$.
If we take the thermodynamic limit and the long-time limit, from the scaling of boson density in \Cref{eq:nzeta}, we obtain
 \begin{align}
   \lim_{n_{\tot} \to \infty}  \lim_{t \gg 1}  \mathcal{P}(t) = 
     \begin{cases}
         1, \   &|\zeta_{L,\tilde{m}}| > 1 \\
        0, \   &|\zeta_{L,\tilde{m}}| < 1
     \end{cases}. \label{eq:dpjump}
 \end{align}
Unlike $\mathcal{P}_0$ in \Cref{eq:jump}, in these two limits the dynamical polarization displays a quantized jump at $|\zeta_{L,\tilde{m}}| = 1$ when the left eigenmode $\u{\tilde{\psi}}_{L,\tilde{m}}$ with smallest real rapidity becomes delocalized. In particular,  from \Cref{eq:ntb} and \Cref{eq:nts}, $|\zeta_{L,\tilde{m}}| = |r|^{-1}$ if $\Delta_s \le 0$ and $|\zeta_{L,\tilde{m}}| = |r_L|$ if $\Delta_s > 0$. $\mathcal{P}(t)$ alone cannot predict gapless phase transitions ($|r^*_Lr_R| = 1$).

Yet, through $\mathcal{P} (t)$, the dynamical process of the boundary-mode separation is made transparent. 
In \Cref{fig:model}~(c), we trace the dynamical polarization in time in the distillation scenario with and without sublattice loss. When $\eta_0 = 0$,  $\mathcal{P}(t)$ evolves from $1/2$  (the center of the chain) to $1$ (the right end), reflecting the  localization direction of dominant NH bulk modes ($|r|^{-1} > 1$). Whereas, when $\eta_0 = 0.4$, $\Delta_s > 0$ [\Cref{fig:model}~(a)]. NH boundary mode localized towards the left end of the chain ($|r_L| < 1$) governs the long-time behaviors of polarization: $\mathcal{P}(t \gg 30) \to 0$.
Interestingly, at short times $0 < t \lesssim 13$, $\mathcal{P}(t)$ keeps increasing under the influence of a larger number of bulk modes, and decreases after $t \gtrsim 13$ when the majority of bulk modes are damped by sublattice loss.

It is natural to define the \emph{polarization drift} to quantify the difference in long-time polarization with and without boundary-mode separation:
   \begin{align}
  D_p  &=  \left.\mathcal{P} (t_\text{max})\right|_{|\eta_0| > |\eta_0|_\text{min}} - \left.\mathcal{P} (t_\text{max})\right|_{\eta_0 = 0}, \label{eq:dyp}
  \end{align}
  where $t_\text{max} \gg 1$.
Illustrated in \Cref{fig:model}~(c), ignoring phase oscillations, $D_p$ is relatively a constant beyond short times. From our former analysis, 
the amplitude of  $D_p$ depends on three factors: $n_\tot$, $|r|^{-1}$ and $|r_L|$.
 When the total length of the chain is finite, the emergence of NH boundary mode is signaled by 
\begin{gather}
    D_p \ne 0, \label{eq:dpf}
\end{gather}
since $|r_L| \ne |r|^{-1}$ in gapped phases. To verify, 
we look at the opposite case $|r_L| = |r|^{-1}$. Applying the relation $r = \sqrt{r_R/r^*_L}$ from \Cref{eq:rRL} and \Cref{eq:r}, one immediately recovers gapless phase transition lines:  $|r_L^*r_R| = 1$, which is consistent with the fact that when NH boundary mode enters the bulk spectrum, it shares the same localization parameters with bulk modes. 
Hence, the polarization drift vanishes at phase transitions: 
\begin{gather}
    \left. D_p \right|_{|r_L^*r_R| = 1} = 0. \label{eq:dpz}
\end{gather}

Next, we illuminate two general features of $D_p$ in \Cref{eq:dpf} and \Cref{eq:dpz} in examples of NH SSH chains with odd and even sites.

\subsubsection{Odd sites}
\begin{figure}[t]
\centering
\includegraphics[width=0.85\columnwidth]{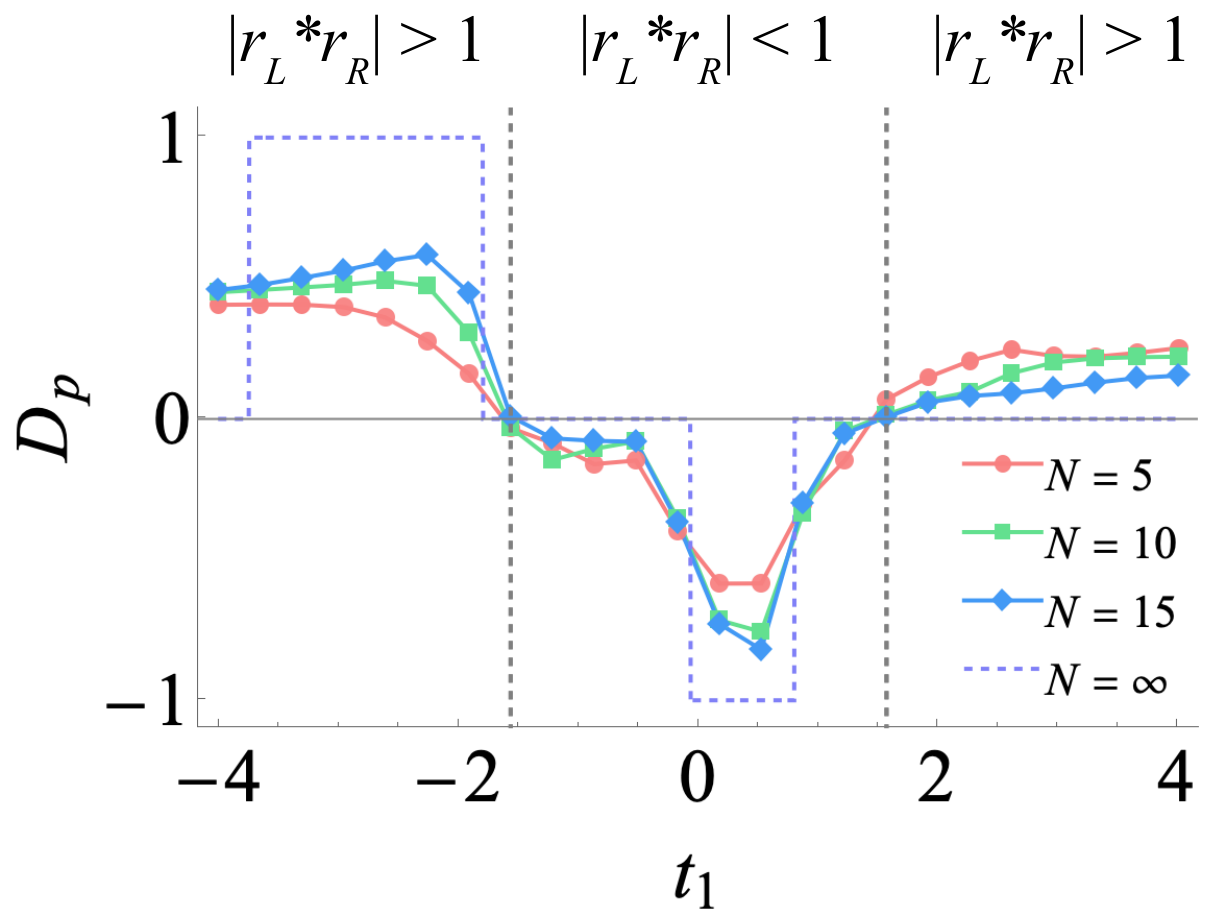}
\caption{{\bf Odd sites: polarization drift $D_p$ alters sign across gapless phase transitions.} 
We choose $n_\text{tot} = 2N-1$ and fix $t_2 =1.5, \eta_1 = 0.5, \eta_2 = 0.2$. 
The dynamical polarization is measured at $t_\text{max}=40$ with  $\eta_0 = 0, 1$.
Phase transitions (vertical dashed lines) occurring at $|r_L^*r_R| = 1$ or  $t_{1,c} = \pm \sqrt{t_2^2-\eta_2^2+\eta_1^2} \simeq \pm 1.568$, are pinpointed by $D_p = 0$ given a finite system size $N$. 
In the thermodynamic limit, $D_p$ converges to the analytical bound (dashed in purple) predicted by \Cref{eq:th}.}
\label{fig:drift}
\end{figure}
The odd-site NH SSH chain has one exact zero-energy boundary mode [\Cref{eq:bmm}] in all gapped phases:  $|r_L^*r_R| < 1$ and $|r_L^*r_R| > 1$. In \Cref{fig:drift},  using our exact solutions, we observe nonzero $D_p$ in these gapped regions as a direct signal of the boundary mode.
When the system size is finite but large enough \textcolor{black}{(e.g. $N = 10$ is already sufficient)} 
$D_p$ drops to zero precisely at phase transitions and alters sign across them: 
\begin{gather}
  \sgn( \left. D_p \right|_{|r_L^*r_R| < 1}) = - \sgn( \left. D_p \right|_{|r_L^*r_R| > 1}). \label{eq:ss}
\end{gather}

In addition, 
 $D_p = 0$ also 
responds to two of the exceptional points (EPs) of $H_{\text{NH}}$: $t_1^+ = 0$ or $t_2^-=0$, rendering 
$|r_L| = |r|^{-1} =  0$ or $\infty$. Yet, these EPs have an order proportional to the system size and \textcolor{black}{are} unstable to any perturbation 
In \Cref{fig:drift}, we sample $t_1 \simeq -0.52$ close to such an EP:   $t_{1, \text{EP}} = -\eta_1 = -0.5$. 
While $D_p \to 0$ as $N$ increases, $D_p$ retains a consistent sign across the EP,
further distinguishing gapless phases transitions from EP crossings.

It is interesting to address the thermodynamic limit where the dynamical polarization becomes an integer in \Cref{eq:dpjump}. One derives a  quantized bound for the polarization drift
(\Cref{fig:drift}, purple dashed curve):
 \begin{gather}
     \left. D_p  \right|_{N \to \infty}{=} \frac{1}{2} \left[ \sgn({\log |r_L|}) - \sgn({\log |r|^{-1}}) \right]. \label{eq:th}
 \end{gather}
It captures $\left. D_p  \right|_{N \to \infty}= \pm 1 $ for $\sgn({\log |r_L|}) \ne \sgn({\log |r|^{-1}})$ and $0$, otherwise. Since 
$\left. D_p  \right|_{N \to \infty}$ vanishes as soon as bulk and boundary modes are localized towards the same direction, it fails to pinpoint critical phase transition lines (\Cref{fig:drift}, vertical gray dashed).

To conclude, enabled by boundary-mode separation, the sign switch of $D_p$ [\Cref{eq:ss}] provides a sharp and robust dynamical signature of phase transitions in finite-size odd-length NH SSH chains.

\subsubsection{Even sites}
\begin{figure}[t]
\centering
\includegraphics[width=0.85\columnwidth]{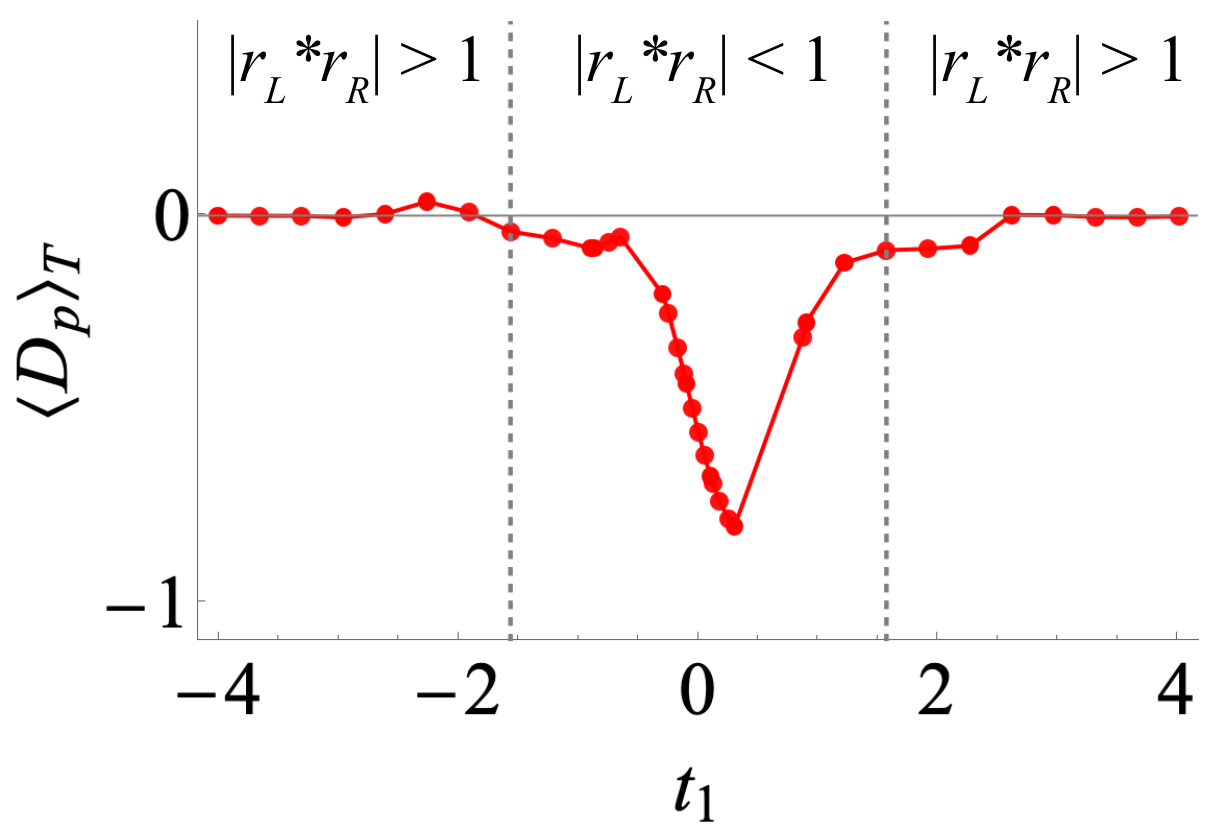}
\caption{{\bf Even sites: polarization drift $D_p$ signals NH topological boundary mode deep inside the topological phase.}
We choose even sites $n_{\text{tot}} = 2N = 30$ and keep parameters $\{t_2, \eta_1, \eta_2, \eta_0 \}$ same as \Cref{fig:drift}. By varying $t_1$, $D_p$ deviates far from zero when $|r_L^*r_R| \ll 1$, elsewhere almost vanishes. To reduce increased phase oscillations in the long-time dynamical polarization, we perform a time-averaged measurement: $\langle D_p \rangle_T = (1/T)\sum_{k=0}^{T-1} D_p(t_\text{max}+k)$ with $t_\text{max}=40$ and $T=20$.}
\label{fig:even}
\end{figure}

The even-site NH SSH chain holds two topological NH boundary modes in the topological gapped phase ($|r_L^*r_R| < 1$) and no boundary mode in the trivial one ($|r_L^*r_R| > 1$). Due to the lack of exact solutions, 
 we provide numerical results on the even-length chain. 
 \Cref{fig:even} shows the behavior of the time-averaged polarization drift in the distillation scenario for $n_\tot = 2N = 30$.
 
 Deep inside the topological region ($|r_L^*r_R| \ll 1$), $D_p$ deviates from zero:
 \begin{gather}
     \left. D_p \right|_{|r_L^*r_R| \ll 1} \ne 0. \label{eq:deep}
\end{gather}
It indicates a successful separation of the emergent topological NH boundary mode exponentially localized on the site $(1,A)$, that is, on the sublattice site $A$ in the first unit cell.
The other boundary mode localized on the site $(N,B)$ is damped by the $B$-sublattice loss  [\Cref{eq:sdb}].
 Far away from the topological region ($|r_L^*r_R| \gg 1$), $D_p$ vanishes in absence of any boundary mode. 
 
Yet, near topological phase transitions ($|r_L^*r_R| = 1$), $D_p$ fluctuates around zero, because of \textcolor{black}{enhanced} finite-size effects.
\textcolor{black}{For the even-length NH SSH chain,} \textcolor{black}{when the system size is small its topological region hosting zero-energy boundary modes} appears narrower than $|r_L^*r_R| < 1$~\cite{flore2018}.  
\textcolor{black}{Moreover, both the eigen energy and the wavefunction of its targeted topological boundary mode localized on the site $(1,A)$ deviate from the exact form in the odd-length chain [\Cref{eq:bmm}]. Analogous to the Hermitian case~\cite{emil2023}, this boundary mode now has a finite non-negligible weight on neighboring $B$ sites with its lifetime reduced by the $B$-sublattice loss.}


In the thermodynamic limit, we envision that in the topological phase, $D_p$ of the even-length chain will converge to the bound in \Cref{fig:drift} ( purple dashed curve) while in the trivial phase,  vanish completely:
  \begin{align}
     &\left. D_p  \right|_{N \to \infty} \\
     &{=}
     \begin{cases}
         \frac{1}{2} \left[ \sgn({\log |r_L|}) - \sgn({\log |r|^{-1}}) \right], & |r_L^*r_R| < 1 \\
        0, & |r_L^*r_R| > 1
     \end{cases}. \notag
 \end{align}

For a finite-size even-length NH SSH chain, the observation that $D_p$ fluctuates across phase transitions ($|r_L^*r_R| = 1$) obscures its role as a topological marker. But a nonvanishing $D_p$ deep inside the topological phase [\Cref{eq:deep}] serves as a good indicator of topological NH boundary mode.



\section{Instabilities and singularities}
\label{sec:is}
In this section, we study the instabilities and singularities of our QBLs in presence of a hybrid bath. For simplicity,  on the Hermitian odd-length SSH chain [\Cref{eq:her}], we introduce coherent bond loss and gain  [\Cref{eq:bs} and \Cref{eq:bsb}] and suppress the sublattice dissipation [\Cref{eq:sdb} and \Cref{eq:sda}]:
\begin{gather}
    |\gamma_0^l|=|\gamma_0^g|=0, \notag \\
 L= |\gamma_1^l| + |\gamma_2^l|,  \quad 
 G= |\gamma_1^g| + |\gamma_2^g|. \label{eq:tlg}
\end{gather}
From the simplified Lyapunov equation of \Cref{eq:rss0}, 
 a formal solution to the covariance matrix $C_s$ is first derived, which if exists will lead to a unique BRSS. Singularities appear when total gain and loss are in balance ($L = G$), while instabilities arise once total gain surpasses loss ($L < G$). Along the solvable lines where the BRSS acquires a uniform density profile, 
 we further study nonequilibrium dynamics of QBLs close to singularities, going from the stable to unstable regions.
 
\subsection{Bosonic reference steady state}

\subsubsection{Existence and uniqueness condition}

It is known that the solution to the Lyapunov equation [\Cref{eq:lya}] exists and is unique if there is no pair of eigenvalues $\bar{\beta}$ of the Liouvillian [\Cref{eq:liou}] that renders ${\bar{\beta}}_m+\bar{\beta}_{m'} = 0, \forall m, m'$ \cite{prosen2010q}.
In terms of the simplified Lyapunov equation 
[\Cref{eq:rss0}], in the following we show a similar constraint imposed on rapidities $\beta$ or eigenvalues of NH damping matrix. 

In \Cref{eq:rss0}, let us apply the decomposition of the damping matrix [\Cref{eq:xdom}] and 
multiply from its  left  and right sides with $\langle \u{\psi}_{Rm'}|$  and $| \u{\psi}_{Rm}\rangle$ respectively. Taking into account the biorthogonal relations $\langle \u{\psi}_{L,m}  | \u{\psi}_{R,m'} \rangle = \delta_{m,m'}$, one gets
\begin{gather}
    (\beta_{m}+\beta_{m'}^*) \langle \u{\psi}_{Rm'}| C_s | \u{\psi}_{Rm} \rangle = \langle \u{\psi}_{Rm'}| M^g | \u{\psi}_{Rm} \rangle. \label{eq:rss1}
\end{gather}
If the rapidities satisfy
 \begin{gather}
 {\beta}_m + {\beta}^*_{m'} \ne 0,  \quad \forall m, m', \label{eq:exist}
 \end{gather}
there exists a unique solution to $C_s$:
 \begin{gather}
     C_{s} =  \sum_{m,m'}  \frac{1}{({\beta}_m+{\beta}^*_{m'})} |\u{{\psi}}_{Lm'} \rangle \langle \u{\psi}_{Rm'}| M^g | \u{\psi}_{Rm} \rangle  \langle \u{{\psi}}_{Lm}|. \label{eq:fs}
 \end{gather}
On the left side of \Cref{eq:rss1}, we have applied the identity operator twice: $\mathbbm{1}_{n\times n} = \sum_m |\u{\psi}_{Rm} \rangle \langle \u{\psi}_{Lm}|$.

We find an important counterexample in our model which violates the existence and uniqueness condition of $C_s$ in \Cref{eq:exist}.
It occurs when the total loss and gain on $t_1$ and $t_2$ bonds become equal. Let us examine the eigenvalues of 
the damping matrix [\Cref{eq:damp0}] in this case:
\begin{gather}
 X  =  iH_\text{S}, \quad \beta_m =  i\epsilon_m, \quad \text{at} \  L = G. \label{eq:sing0} 
\end{gather}
From combined $P$ and $K$ symmetries [\Cref{eq:psym} and \Cref{eq:ksym}], the eigenvalues  of $H_\text{S}$ obey: $\{ \epsilon_m \} = \{ -\epsilon_m \} = \{ -\epsilon_m^{*} \} =  \{ \epsilon_m^{*} \}$.
One can always find one pair of rapidities that invalidates the formal solution to $C_s$ in \Cref{eq:fs}:
\begin{gather}
 \beta_m + \beta^*_{m'} =  i (\epsilon_m - \epsilon_{m'}^*) = 0, \quad \exists m,m'. \label{eq:sing}
\end{gather}
Since our argument is based on symmetries, one can generalize $H_\text{S}$ in \Cref{eq:sing0} to generic $PK$-symmetric NH Hamiltonians, for which the singularities of underlying QBLs arise on the global-balance line ($L=G$).

We can also treat our Hamiltonian $H_\text{S}$, the odd-length NH SSH chain in \Cref{eq:hs0} in more detail. \Cref{eq:rss1} is valid for any pair of rapidities. Along the global-balance line, the gain matrix $M^g$ given by \Cref{eq:mg} is nonzero. In \Cref{eq:rss1},
 one can then take the exact zero-energy boundary eigenmode [\Cref{eq:bmm}]: $m=m'=0$ , such that $\beta_0 + \beta^*_0 = 0$.   $C_s$  has no solution because $\langle \u{\psi}_{R0}| M^g | \u{\psi}_{R0} \rangle$ is finite.

\subsubsection{Solvable lines towards a uniform structure}
\begin{figure}[t]
\centering
\includegraphics[width=0.85\columnwidth]{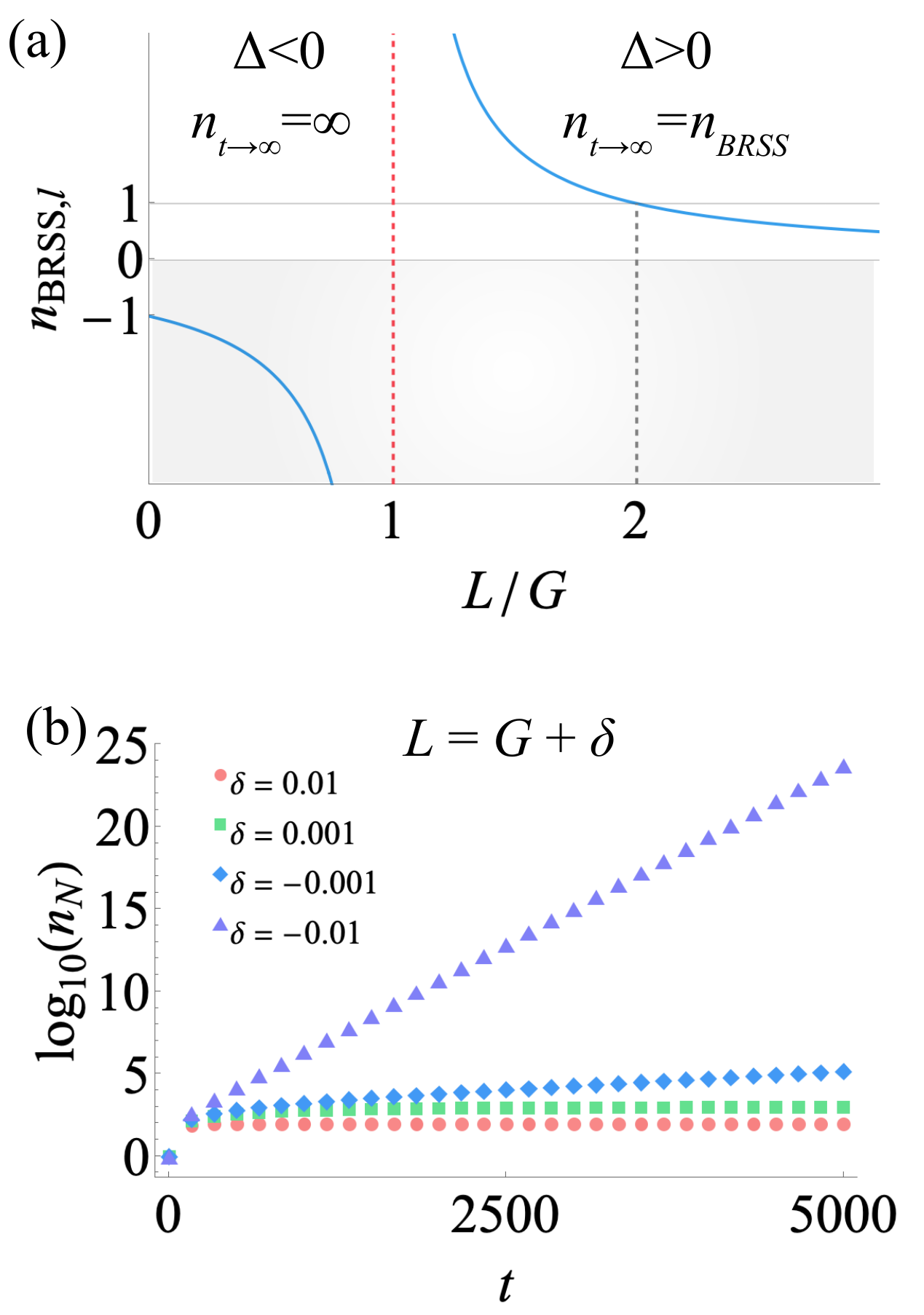}
\caption{{\bf Stability of bosonic reference steady state.}  For the bond-dissipative SSH chain, we focus on the solvable lines of \Cref{eq:sol}. In (a),  we show the boson density of a uniform BRSS as a function of the rate between total bond loss ($L$) and bond gain ($G$).  With a positive Liouvillian gap, the system can be stabilized to the BRSS at long times. Otherwise, the system becomes unstable with diverging boson density.  (b) Boson density as a function of time close to the global-balance singular line: $L=G+\delta$, $\delta \to 0^{\pm}$. We take $|\gamma_1^l|= 1 + \delta$, $|\gamma_1^g| =1 $, $|\gamma_2^l| = |\gamma_2^g|=0$. It leads to $\eta_1 = \frac{1}{2}\delta$, $\gamma_1= 1+\frac{1}{2}\delta$,  $\eta_2 = \gamma_2 = 0$. Fulfilling $\gamma_1\gamma_2 = 0$, the BRSS exhibits a uniform density profile.
We further fix
$t_1 = 1$, $t_2 = 2$ so that $\Delta =2(\eta_1 + \eta_2) = \delta$.  Choosing a system size $n_\text{tot} = 2N-1 = 29$, the boson density is measured at the middle of the chain: $n_l (t) = n_N (t)$. At $t=0$, the initial state holds one boson per site. As time advances, for $\delta = \Delta \to 0^+$, the system relaxes to the BRSS: $n_{t \to \infty, N}  = n_{\text{BRSS}, N}  = G/\delta = 10^2, 10^3$ at $\delta = 0.01, 0.001$. Whereas, for $\delta = \Delta  \to 0^-$, we find $n_N(t) = c_0 e^{|\delta|t}$ when $t \gg 1$.}
\label{fig:brss}
\end{figure}

Away from $L=G$, the global-balance singular line,
one can find solvable lines for $L\neq G$ along which $C_s$ holds a simple uniform structure.
Assuming 
 \begin{gather}
     C_s = a \cdot \mathbbm{1}_{n \times n} \label{eq:us} 
 \end{gather}
 in the simplified Lyapunov equation of \Cref{eq:rss0} and applying  $M^g$ and $X$ from \Cref{eq:mg} and \Cref{eq:damp0}, we obtain 
 \begin{align}
    (|\gamma_1^l|-|\gamma_1^g|) a &= |\gamma_1^g|, \notag \\
    (|\gamma_2^l|-|\gamma_2^g|) a &=|\gamma_2^g|, \label{eq:a}
 \end{align}
Recalling and introducing auxiliary variables 
  \begin{gather}
\eta_i = \frac{1}{2}(|\gamma_i^l| -|\gamma_i^g|),  \quad \gamma_i = \frac{1}{2}(|\gamma_i^l| + |\gamma_i^g|),
      \label{eq:geta}
  \end{gather}
$C_s$ in \Cref{eq:us} is solvable if 
  \begin{gather}
    \frac{\eta_1}{\gamma_1} = \frac{\eta_2}{\gamma_2} \ne 0 \quad \text{or} \quad \gamma_1 \gamma_2 = 0.   \label{eq:sol}
  \end{gather}
It renders a uniform density profile for the BRSS ($n_{\text{BRSS},l} = C_{s, ll} = a, \forall l$):
 \begin{gather} 
   n_{\text{BRSS},l} =  \frac{G}{L-G} \in (-\infty, \infty). \label{eq:nbrss}
 \end{gather}
 By definition [\Cref{eq:tlg}], $L \ge 0, G \ge 0$ which means $n_{\text{BRSS},l}$ is unbounded.  Illustrated in \Cref{fig:brss}~(a), we can recover several interesting limits. In the case of a pure bath with only bond loss (gain), $n_{\text{BRSS},l} = 0$ at $G=0$ ($-1$ at $L=0$), consistent with our earlier results in \Cref{eq:cs0} and \Cref{eq:cs1}. Approaching the global-balance singular line at $L/G = 1$ [\Cref{fig:brss}, red dashed], the boson density of the BRSS diverges: $n_{\text{BRSS},l} \to \pm \infty$, in agreement with the fact that a unique solution to  $C_s$ does not exist.

\subsubsection{Stabilizing BRSS with a positive Liouvillian gap}
 From \Cref{sec:lsg}, we learn that it requires a positive Liouvillian gap for the BRSS to become the steady state at long times,  for which reason the Liouvillian gap is also referred to as the stability gap of QBLs \cite{viola2024}. In absence of sublattice dissipation, $\Delta_s = \Delta_\text{bulk} -  \Delta_\text{boundary} \le 0$. The Liouvillian gap of our system is determined by bulk modes: $\Delta = \Delta_\text{bulk}$. From \Cref{eq:dbulk}, it is clear that 
 \begin{gather}
     \sgn [\Delta] =  \sgn[\eta_1 + \eta_2] = \sgn[L-G]. 
 \end{gather}
 
When the total loss on bonds is larger than gain, $\Delta > 0$, the BRSS in \Cref{eq:nbrss} can be stabilized at long times: $n_{t \to \infty,l} = n_{\text{BRSS},j},\forall l$. In particular, as shown in \Cref{fig:brss}~(a), to restrain the steady-state boson density $n_{t \to \infty,l}  \in [0, 1)$, it requires $L/G > 2$. Otherwise, $n_{t \to \infty,l} > 1$ for intermediate bond loss $1 < L/G  < 2$.

When the bond gain surpasses loss, $\Delta < 0$. The system becomes unstable with boson densities diverging beyond short times: $n_{t \to \infty,l} = \infty, \forall l$. Depicted by blue curves in \Cref{fig:brss}~(a), $n_{\text{BRSS},l}$ simultaneously acquires an unphysical  negative value as soon as $L/G < 1$.

 It is useful to compare with quadratic fermionic Lindbladians, of which the Liouville space is bounded. In a similar bond-dissipative quantum setup~\cite{yang2022}, under solvable conditions, the  fermionic steady state (FSS) can be endowed with the following uniform density profile ($\forall l$):
\begin{gather}
 n_{\text{FSS},l}= \frac{G}{L+G} \in [0,1].
 \end{gather}
Here, the Pauli exclusion principle is naturally respected. Under OBC, all the rapidities have positive real parts ensuring $\Delta > 0$. The FSS can always be reached at long times regardless of relative strengths of bond loss and gain. Along the global-balance line ($L=G$), singularities also disappear with the fermionic chain stabilized at the half-filling steady state: $n_{\text{FSS},l}=1/2, \forall l$.

\subsection{Nonequilibrium dynamics close to singularities}
In the end, we perturb our QBLs around singularities and study emergent dynamics in terms of nonequilibrium density evolution. \textcolor{black}{It turns out that the long-time boson density is extremely sensitive to any small imbalance between loss and gain around singularities [\Cref{fig:brss}~(b)]. It provides insights for future experiments to locate potential unknown singularities in QBLs and helps set the boundaries between stable and unstable regimes.} 

It is convenient to apply solvable lines in \Cref{eq:sol} associated with a uniform BRSS. We look at following limits: 
\begin{gather}
    L-G  = \delta  \to 0^\pm,
\end{gather}
which along solvable lines renders a covariance matrix $C_s = (G/\delta) \cdot \mathbbm{1}$.
Choosing the initial state to be one boson per site: $C(0) = \mathbbm{1}$, from the two-point correlator in \Cref{eq:tc}, the boson density  evolves as 
    \begin{gather}
  n_l(t) = \frac{1}{\delta} \left[G + (\delta-G) Q(l,t)\right], \label{eq:den1}
\end{gather}
where the propagator is given by \Cref{eq:prop}. In absence of sublattice dissipation, the Liouvillian gap is determined by bulk modes in \Cref{eq:dbulk}. Suppose $|\eta_1| < |t_1|$, $|\eta_2| < |t_2|$, one finds
\begin{gather}
    \Delta = 2(\eta_1 + \eta_2) = \delta.
\end{gather}
Shown in \Cref{fig:brss}~(b), when $\delta =\Delta  \to 0^+$, as time advances the system evolves to the BRSS with uniform density: 
\begin{gather}
   n_{t \to \infty, l} =  n_{\text{BRSS},l} = \frac{G}{\delta}, \quad \forall l.
\end{gather}
As soon as $\delta = \Delta  \to 0^-$, the system becomes unstable with boson density exponentially increasing with time:
\begin{gather}
n_{t \to \infty, l} = \left. c_0e^{vt}\right|_{t \to \infty}= \infty, \quad \forall l.
\end{gather}
From the propagator in \Cref{eq:prop}, the growth rate is controlled by the Liouvillian gap, or the imbalance strength of total bond loss and gain: 
\begin{gather}
   v \simeq -\Delta = |\delta|. \label{eq:speed}
\end{gather}
In \Cref{fig:brss}~(b), we perform a fitting analysis in the long-time domain $t \in [2500, 5000]$. The exponential growth rate is found to be $v  \simeq 0.0010288$ at $\delta = -0.001$, and  $v  \simeq 0.009998$ at $\delta = -0.01$, in agreement with our prediction.

\section{Discussion}
To summarize, we have introduced a full open  quantum framework where  engineered dissipation enables NH boundary modes to be generated, separated, and utilized to  dynamically characterize NH topology. These boundary modes arise from the interplay between  a parent Hermitian Hamiltonian and coherent bond dissipation. With additional sublattice dissipation, 
the distillation and amplification schemes presented realize both \textcolor{black}{the Liouvillian skin effects of NH boundary modes}  (\Cref{fig:bm}) 
and a direct, measurable signature of NH topology through polarization drift (\Cref{fig:drift} and \Cref{fig:even}).

Our work invites several theoretical generalizations. 
When the sample and bath are not clean, analogous to its Hermitian counterpart~\cite{emil2023}, the exact NH boundary mode [\Cref{eq:bmm}] remains robust against disorders on the  bonds ($t_1$, $t_2$) and on the bond dissipation ($\eta_1$, $\eta_2$). 
Our framework can also be extended to higher-dimensional and multi-band NH systems, including those with broken $P$ symmetry. In particular, high-order NH boundary modes~\cite{elisabet2019,yang2024a} once dynamically separated will bring enriched anomalous quantum dynamics with the current flow driven towards corners, edges, or surfaces of nonreciprocal lattices.  Moreover, with the recent discovery of inducing the multifractality in open systems through the Cayley tree graph \cite{hamanaka2025}, our protocol can be applied to unveil hidden multifractal statistics  of NH boundary modes.

Another avenue is the refinement of dissipation schemes: replacing sublattice loss with single-site loss~\cite{emil2023} is still compatible with weak symmetries in the late-time mixed states~\cite{prosen2012n,diehl2025t}, potentially broadening applicability from non-interacting to interacting many-body systems~\cite{wang2024n,marco2025}.


\section*{Acknowledgements}
We thank Oscar Arandes, Christopher Ekman and Hui Liu for discussions. 
This work was supported by the Knut and Alice Wallenberg Foundation (KAW) via the Wallenberg Academy Scholars program (2023.0256) and the project Dynamic Quantum Matter (2019.0068), the Swedish Research Council under the VR Starting Grant 2024-05213, and the G\"oran Gustafsson Foundation for Research in Natural Sciences and Medicine.

\renewcommand{\theequation}{\Alph{section}\arabic{equation}}
\appendix
\section{\uppercase{{NH damping matrix of QBLs}}}

In the Appendix, we present two equivalent methods of deriving NH damping matrix of QBLs in \Cref{eq:x}.
 In the first approach, we apply third quantization over unbounded Liouville space of bosons and diagonalize the quadratic Lindbladian. In the second approach, we solve the time evolution of two-point correlation function through commutation relations of complex bosons.

\subsection{Third quantization}
Third quantization has been generalized to solve Lindblad master equation in open bosonic systems \cite{prosen2010q} with recently proposed reformulations \cite{barthel2022} and connections to the Keldysh field-theory approach \cite{clerk2023t}. Here, we give a brief review of the original formalism \cite{prosen2010q} and draw a comparison with open Fermi matter \cite{prosen2008,prosen2010ex,prosen2010sp,yang2022}.

Let us denote the unbounded Liouville space of bosons as $\mathcal{K}$. To find its Fock basis, given a density matrix $\rho \in \mathcal{K}$, we introduce left and right multiplication maps for each bosonic operator on the $j$-th site: $b_j^L \rho = b_j\rho$, $b_j^{\dagger L} \rho = b_j^\dagger\rho$ and $b_j^R \rho = \rho b_j$, $b_j^{\dagger R} \rho = \rho b_j^\dagger$. Two sets of creation and annihilation operators $b_{\nu}, b'_{\nu}$ with $\nu \in \{0,1\}$ can be built according to
 \begin{gather}
     b_{0,j} = b_j^L, \quad b'_{0,j} = b_j^{\dagger L} - b_j^{\dagger R}, \notag \\  
     b_{1,j} = b_j^{\dagger R}, \quad 
     b'_{1,j} = b_j^{R} - b_j^{L}, \label{eq:b01}
 \end{gather}
such that they satisfy commutation relations of bosons over the space $\mathcal{K}$:
  \begin{gather}
      [b_{\nu,j}, b'_{\mu, k}] = \delta_{\nu,\mu} \delta_{j,k}, \  [b_{\nu,j}, b_{\mu, k}] = [b'_{\nu,j}, b'_{\mu, k}] = 0.
  \end{gather}
  While the vacuum of $\mathcal{K}$ is a pure state $\rho_0 = |\psi_0\rangle \langle \psi_0|$ determined by annihilation operators $b_\nu$:
  \begin{gather}
     b_{\nu,j} \rho_0 =0, \quad \forall \nu, \ \forall j,  
  \end{gather}
the Fock basis $|\u{\alpha} \rangle$ is unbounded constructed from creation operators $b'_\nu$ acting on the vacuum:
\begin{gather}
     \quad  |\u{\alpha} \rangle = \prod_{\nu,j} \frac{(b'_{\nu,j})^{\alpha_{\nu, j}}}{\sqrt{\alpha_{\nu,j}!}} \rho_0, \quad \alpha_{\nu, j} \in \mathbb{Z}_+. \label{eq:bn}
\end{gather}
Here, $\u{\alpha} = (\alpha_{0,1}, \alpha_{1,1},  \dots, \alpha_{0,n}, \alpha_{1,n})^T$ consists of $2n$ components for a system of $n$ sites. Notably, the boson number is unbounded: $\alpha_{\nu, j} \in \{0, 1,2,3, \dots \}$. In comparison, for open Fermi matter, the Liouville space is expanded by a basis of real Majorana fermions,  of which the occupation number is restricted by the Pauli exclusion principle: $\alpha_j \in \{0,1\}$.

Applying the left and right multiplication maps to Lindblad master equation in \Cref{eq:lin}, the Liouvillian shares the form: 
\begin{gather}
    \hat{\mathcal{L}} = -i (\mathcal{H}^L -\mathcal{H}^R)  + \sum_{\mu} [ L_\mu^L L_\mu^{\dagger R}
     - \frac{1}{2}( L_\mu^{\dagger L} L_\mu^L + L_\mu^R L_\mu^{\dagger R} )],
\end{gather}
with the ingredients,
\begin{gather}
     \mathcal{H} = \u{b}^\dagger \cdot H \u{b}, 
     \quad L^l_\mu = \u{l}_\mu \cdot \u{b}, \quad L^g_\mu = \u{g}_\mu \cdot \u{b}^\dagger. \label{eq:hlg}
 \end{gather}
The mapping in \Cref{eq:b01} enables us to change to a natural basis $\u{c} = (\u{a}, \u{a}')^T = (\u{b}_0, \u{b}_1, \u{b}'_0, \u{b}'_1)^T$, and arrive at a quadratic Liouvillian:
\begin{gather}
    \hat{\mathcal{L}} = \u{c}^T \cdot S \u{c} + S_0 \cdot \mathbbm{1}_{4n \times 4n}. \label{eq:lv}
\end{gather} 
The $4n \times 4n$ matrix $S$  takes the form:
 \begin{gather}
     S = \begin{pmatrix}
         0 & -\bar{X} \\
         -\bar{X}^T & Y 
     \end{pmatrix}, \label{eq:edamp} 
     \end{gather}
     where
     \begin{gather}
         \bar{X} = \frac{1}{2} 
     \begin{pmatrix}
      X & 0 \\
      0 & X^*
     \end{pmatrix}, \quad  Y = \frac{1}{2} 
     \begin{pmatrix}
      0 & (M^g)^T \\
      M^g & 0
     \end{pmatrix}. \label{eq:xy}
 \end{gather}
Its elements consist of the $n \times n$ NH damping matrix:
 \begin{gather}
     X = iH^T + \frac{1}{2}[(M^l)^T -M^g], \label{eq:damp} 
 \end{gather}
with contributions from loss and gain matrices:
 \begin{gather}
     M^l_{ij} = \sum_\mu \u{l}_{\mu}^*(i) \u{l}_{\mu}(j), \quad M^g_{ij} = \sum_\mu \u{g}_{\mu}^*(i) \u{g}_{\mu}(j). \label{eq:mlg}
 \end{gather}
The scalar in the Liouvillian of \Cref{eq:lv} takes the value $S_0 =  \tr \bar{X}  = \frac{1}{2}[\tr(M^l) - \tr(M^g)]$. 

Next, we show the bosonic Liouvillian can be diagonalized via the same set of eigenmodes of the enlarged damping matrix $\bar{X}$ in \Cref{eq:edamp}, if there exists a unique solution to 
the continuous Lyapunov equation:
\begin{gather}
   \bar{X}^T Z + Z \bar{X} = Y. \label{eq:lya}
\end{gather}
Slightly different from Ref.~\cite{prosen2010q},
we give a proof based on transformation matrices.
If the Lyapunov equation in \Cref{eq:lya} is satisfied, the Liouvillian shares a block-diagonalized form: 
 \begin{gather}
     S = J^{-1} W \begin{pmatrix} 
      -\bar{X}^T & 0 \\
      0 & \bar{X}
     \end{pmatrix} W^{-1},  \label{eq:tm}
 \end{gather}
 under two transformation matrices:
  \begin{gather}
      J = \begin{pmatrix}
      0 & 1 \\
      -1 & 0
      \end{pmatrix},  \quad
      W =\begin{pmatrix}
        1 & Z \\
        0 & 1
      \end{pmatrix}. \label{eq:tm1}
  \end{gather} 
More precisely, 
\begin{gather}
      \hat{\mathcal{L}} = \begin{pmatrix}
      \u{\upsilon}'^T & -\u{\upsilon}^T \end{pmatrix} \cdot  \begin{pmatrix} 
      -\bar{X}^T & 0 \\
      0 & \bar{X}
     \end{pmatrix} \begin{pmatrix}
 \u{\upsilon} \\ \u{\upsilon}'
\end{pmatrix} + \tr \bar{X} \cdot \mathbbm{1},
\end{gather}
where a new basis is identified as 
\begin{gather}
\begin{pmatrix}
 \u{\upsilon} \\ \u{\upsilon}'
\end{pmatrix}=   W^{-1} \u{c} = \begin{pmatrix}
 \u{a} -Z \u{a}'\\ \u{a}'
\end{pmatrix},
\end{gather}
with the commutation relations:
\begin{gather}
 [\upsilon_j, \upsilon'_k] = \delta_{j,k}. \label{eq:up}   
\end{gather}
Suppose the enlarged damping matrix $\bar{X}$ is diagonalizable via a nonunitary  matrix $U$,
\begin{gather}
    \bar{X} = U \Sigma U^{-1}, \quad \Sigma = \diag\{ \bar{\beta}_1, \dots, \bar{\beta}_{2n} \}, \label{eq:tdd}
\end{gather}
one immediately arrives at a decomposition of the Liouvillian in its normal master modes (NMMs):
 \begin{gather}
      \hat{\mathcal{L}} = -2 \sum_{m=1}^{2n} \bar{\beta}_m \u{\zeta}'_m \cdot \u{\zeta}_m. \label{eq:liou}
 \end{gather}
The NMMs are given by
 \begin{gather}
     \u{\zeta} = U^T \u{\upsilon}, \quad \u{\zeta}' = U^{-1} \u{\upsilon}',
 \end{gather}
and also satisfy the commutation relations: 
 \begin{gather}
    [{\zeta}_m, {\zeta}'_l] = \delta_{m,l}.
 \end{gather}
 So far, the spectrum of $\hat{\mathcal{L}}$ incorporates the eigenvalues ${\bar{\beta}} = \{ \bar{\beta}_1, \dots, \bar{\beta}_{2n}\}$ of $\bar{X}$. From \Cref{eq:xy},  ${\bar{\beta}}$ has a group structure of complex conjugate pairs: 
\begin{gather}
 {\bar{\beta}} = \{ \frac{1}{2}\beta, \  \frac{1}{2}\beta^* \}.
\end{gather}
 ${{\beta}} = \{ {\beta}_1, \dots, {\beta}_{n}\}$ collects $n$ eigenvalues of the damping matrix  $X$. Combined with the extra factor $2$ in \Cref{eq:liou}, hereafter we refer to ${{\beta}}$ as rapidity spectrum.  

In the end, as the solution to the Lyapunov equation, the matrix $Z$ encodes all the information on the covariance matrix $C_s$.
Applying to \Cref{eq:lya} the relation
 \begin{gather}
     Z = \begin{pmatrix}
      0 & C_s^* \\
      C_s^\dagger & 0
     \end{pmatrix}, \label{eq:z}
 \end{gather}
 we recover the definition of the covariance matrix: $\partial_t C_s = 0$ in \Cref{eq:rss0}.
The covariance matrix $C_s$ is widely used as a reference point in calculating dynamical  observables, e.g. $\tilde{C}(t) = C(t) - C_s$ in \Cref{eq:tc}. In a physical picture, 
$C_s$  determines the two-point correlator  of a
bosonic reference steady state.
Following the time evolution of the density matrix  $\rho_s = \rho (t)|_{t \to \infty} = e^{\hat{\mathcal{L}t}} \rho_0|_{t \to \infty}$, QBLs can relax to the BRSS given a positive Liouvillian gap $\Delta = 2 \min_{\forall m} \re[\beta_m] > 0$. 
Whereas, any steady state is precluded once the Liouvillian gap  turns negative  or $\exists m, \re[\beta_m] < 0$.

\subsection{Equation of motion of two-point correlator}
We provide an alternative derivation of NH damping matrix through the equation of motion of two-point correlation function.  Freedom of basis choice allows one to apply either a real symmetric basis corresponding to position and momentum operators for bosons \cite{eisert2010}: $u_{1,j} = b_j + b_j^\dagger$ and $u_{2,j} = i(b_j - b_j^\dagger)$, or a complex boson basis \cite{clerk2023t}: $b_j$ and $b_j^\dagger$. The two representations are analogous to   Majorana fermion \cite{alba2021} and complex fermion \cite{fei2019} operators  in open fermionic systems. For simplicity, we adopt the original complex boson basis.

The time evolution of the two-point correlator  $\partial_t C_{jk}(t) = \tr [\partial_t \rho(t) b_j^\dagger b_k]$ can be resolved directly by applying  the Lindblad master equation in \Cref{eq:lin}. Contributions from $\mathcal{H}$, $L_\mu^l$ and $L_\mu^g$ are differentiated as $\partial_t C = A^0 + A^l + A^g$. For each part, we use the cyclic property of the trace and commutation relations of bosons $[b_j,b_k^\dagger] = \delta_{j,k}$. With the notations in \Cref{eq:hlg} and \Cref{eq:mlg}, one obtains
\begin{align}
    A^0_{jk} &= -i \sum_{mn} H_{mn}\tr ([b_m^\dagger b_n, \rho]b_j^\dagger b_k) \notag \\
    &= -i \sum_{mn} H_{mn} \tr (\rho[b_j^\dagger b_k, b_m^\dagger b_n]) \notag \\
    &=  -i \sum_{mn} H_{mn} \tr (\rho (b_j^\dagger [b_k, b_m^\dagger] b_n - b_m^\dagger [b_n, b_j^\dagger]b_k) ) \notag \\
  &= i \sum_m H^T_{jm} C_{mk} -i \sum_n  C_{jn} H_{nk}^T, 
  \end{align}
  \begin{align}
   A^l_{jk} &= \sum_{mn} M_{mn}^l \tr (b_n \rho b_m^\dagger b_j^\dagger b_k - \frac{1}{2} \{b_m^\dagger b_n, \rho \} b_j^\dagger b_k) \notag \\
   &= \sum_{mn} M_{mn}^l \tr (\rho(b_m^\dagger [b_j^\dagger b_k, b_n] + \frac{1}{2}[b_m^\dagger b_n, b_j^\dagger b_k])) \notag \\
  &= \sum_{mn} M_{mn}^l \tr  (\rho(-\frac{1}{2} b_m^\dagger [b_n,b_j^\dagger] b_k - \frac{1}{2}b_j^\dagger[b_k, b_m^\dagger]b_n) )\notag \\
  &=- \frac{1}{2} \sum_m (M^l)^T_{jm} C_{mk} - \frac{1}{2} \sum_n C_{jn}(M^l)^T_{nk}, 
  \end{align}
  \begin{align}
     A^g_{jk} &= \sum_{mn} M_{mn}^g \tr (b_n^\dagger \rho b_m b_j^\dagger b_k - \frac{1}{2} \{b_m b_n^\dagger, \rho \} b_j^\dagger b_k) \notag \\
   &= \sum_{mn} M_{mn}^g \tr (\rho(b_m [b_j^\dagger b_k, b_n^\dagger] - \frac{1}{2}[b_j^\dagger b_k,b_m b_n^\dagger])) \notag \\
  &= \sum_{mn} M_{mn}^g \tr (\rho (\frac{1}{2} b_j^\dagger b_m [b_k,b_n^\dagger]  + \frac{1}{2}[b_m, b_j^\dagger]b_n^\dagger b_k \notag \\
  &\phantom{= \sum_{mn} M_{mn}^g \tr (}+ \delta_{m,j}[b_k,b_n^\dagger]) \notag \\
  &=   \frac{1}{2} \sum_n M^g_{jn} C_{nk} + \frac{1}{2} \sum_m C_{jm}M^g_{mk} + M^g_{jk}.
\end{align}
The equation of motion of two-point correlator follows the compact form: 
\begin{gather}
    \partial_t C(t) = -X^\dagger C(t) - C(t)X + M^g, \notag \\
    X = iH^T + \frac{1}{2}[(M^l)^T - M^g], \label{eq:emo}
\end{gather}
where the Hermiticity of $H$, $M^l$ and $M^g$ is also employed. We thus retrieve the same structure of NH damping matrix as in \Cref{eq:damp}.


\bibliography{sample0}

\end{document}